\documentclass{article}[fulltext]

\usepackage[font=scriptsize]{caption}
\usepackage[utf8]{inputenc} % allow utf-8 input
\usepackage[T1]{fontenc}    % use 8-bit T1 fonts
\usepackage{url}            % simple URL typesetting
\usepackage{booktabs}       % professional-quality tables
\usepackage{amsfonts}       % blackboard math symbols
\usepackage{nicefrac}       % compact symbols for 1/2, etc.
\usepackage{microtype}      % microtypography
\usepackage{xcolor}         % colors
\usepackage{color}
\usepackage{amsmath}
\usepackage{amsthm}
\usepackage{algorithm}
\usepackage[noend]{algpseudocode}
\usepackage{xr}
\usepackage{graphicx}
\usepackage{enumerate}

\makeatletter
\def\BState{\State\hskip-\ALG@thistlm}
\makeatother

\newcommand{\vir}[1]{``#1"}

\newtheorem{theorem}{Theorem}[section]
\newtheorem{corollary}[theorem]{Corollary}

\newtheorem{definition}[theorem]{Definition}
\begin{document}

\title{ From Specific to Generic Learned Sorted Set Dictionaries: A Theoretically Sound Paradigm Yelding Competitive Data  Structural Boosters in Practice}

	\author{Domenico Amato\\
	\and
	Giosu\'e Lo Bosco\\
	\and
	Raffaele Giancarlo}

\date{
	Dipartimento di Matematica e Informatica\\ 
	Universit\'a degli Studi di Palermo, ITALY\\
	%{\{domenico.amato,andrea.desalve,giosue.lobosco,raffaele.giancarlo\}@unipa.it}\\
	\today
}

\maketitle 
	
	\begin{abstract}
		This research concerns Learned Data Structures, a recent area that has emerged at the crossroad of Machine Learning and Classic Data Structures. It is methodologically important and with a high practical impact. We focus on Learned Indexes, i.e., Learned Sorted Set Dictionaries. The proposals available so far are specific in the sense that they can boost, indeed impressively,  the time performance of  Table Search Procedures with a sorted layout only, e.g., Binary Search. We propose a novel paradigm that, complementing known specialized ones, can produce Learned versions of any Sorted Set Dictionary, for instance, Balanced  Binary Search Trees or Binary Search on layouts other that sorted, i.e., Eytzinger. Theoretically, based on it,  we obtain several results of interest,  such as  (a) the first Learned Optimum Binary Search Forest, with mean access time bounded by the Entropy of the probability distribution of the accesses to the Dictionary; (b) the first Learned Sorted Set Dictionary that,  in the Dynamic Case and in an amortized analysis setting,  matches the same time bounds known for Classic Dictionaries. This latter under widely accepted assumptions regarding the size of the Universe. The experimental part, somewhat complex in terms of software development,  clearly indicates the non-obvious finding that the generalization we propose can yield effective and competitive Learned Data Structural Booster, even with respect to specific benchmark models. 
	\end{abstract}

	\section{Introduction}
	
	With the aim of obtaining time/space improvements in classic Data Structures, an emerging trend is to combine Machine Learning techniques with the ones proper for Data Structures. This new area goes under the name of Learned Data Structures. It was initiated by \cite{kraska18case}, it has grown very rapidly \cite{Ferragina:2020book} and now it has been extended to include also Learned  Algorithms \cite{Mitz20}, while the number of Learned Data Structures grows \cite{Boffa21}. In particular, the theme common to those new approaches to Data Structures Design and Engineering is that a query to a data structure is either intermixed with or preceded by a query to a Classifier \cite{duda20} or a Regression Model \cite{FreedmanStat}, those two being the learned part of the data structure.  Learned Bloom Filters \cite{Zhenwei2020,Fumagalli:2021,kraska18case,Mitz18,vaidya2020partitioned}  are an example of the first type, while Learned Indexes, here referred to as Learned Sorted Set Dictionaries,  are examples of the second one  \cite{amato2021lncs,amato2023nn,amato2022standard,amato2021learned,amato22eann,Ferragina:2020book,Kipf19,Kipf20,Marcus20}.  A \vir{predecessor} of those models has been proposed in \cite{Ao11}. Those latter are the object of this research.

	\subsection{Learned Searching in Sorted Sets}
\begin{figure}[tbh]
		%\begin{center}
		\centering
		\includegraphics[scale=0.8]{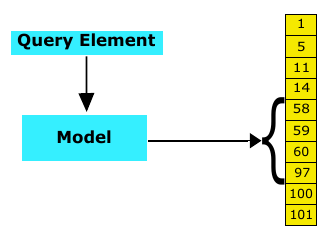}
		\caption{{\bf  A general paradigm of  Learned Searching in a Sorted Set \cite{Marcus20}}. The model is trained on the data in the table. Then, given a query element, it is used  to predict the interval in the table where to search (included in brackets in the figure).}
		%\end{center}
		\label{fig:Par}
	\end{figure}

	With reference to Figure \ref{fig:Par}, a  generic paradigm for learned searching in sorted sets consists of a model, trained over the input data. 
	As described in Section \ref{sec:PS} here and Section \ref{S-sec:models} of the Supplementary File, such a model may be as simple as a straight line or more complex, with a tree-like structure. It is used to make a prediction regarding where a  query element may be in the sorted table.  Then, the search is limited to the interval so identified and, for the sake of exposition, performed via Binary Search.  
	It is quite remarkable then that the novel model proposals,  sustaining Learned Dictionaries,   are quite effective at speeding up Binary Search at an unprecedented scale and are competitive with respect to even more complex Dictionary structures, i.e., {\bf B-Trees} \cite{comer1979ubiquitous}.  Indeed, a recent benchmarking study \cite{Marcus20} (see also \cite{Kipf19}) shows quite well how competitive 
	those Learned Data Structures are.  Another, more recent,  study offers an in-depth analysis of Learned Dictionaries and provides recommendations on when to use them as opposed to other data structures \cite{Mailtry21}. In theoretic terms, those Learned Models yield search procedures that, in the worst-case scenario,  are no worse than the basic routines we have mentioned earlier, provided that the prediction can be made in $O(\log n)$ time, where $n$ denotes the size of the set to be searched into. However,   \emph{de facto} they are boosters of the time performance of Sorted Table Set procedures, limited to  Classic  Binary Search \cite{KnuthS} and Interpolation Search \cite{Peterson57}. Indeed, critical to their proper working is the use of one of those two procedures in the final search stage. Other array layouts for Binary Search, for instance very efficient ones such as Eytzinger \cite{Morin17}, cannot be used by those models. More in general, Sorted Set Dictionaries, other than the two already mentioned, cannot be used. Therefore, the boosting power of those Models is methodologically limited and a natural question is: to what extent Learned Models for Sorted Set Dictionaries can be made \emph{generic}, i.e., able to work for \emph{any} Sorted Set Dictionary handling the final search stage. A related question, to be settled inherently experimentally, is to what extent the boosting effect obtained for Binary and Interpolation Search would apply to relevant classes of Sorted Set Dictionaries, e.g., array layout other than sorted \cite{Morin17}, Balanced Search Trees \cite{Cormer2009}, Cache Oblivious Search Trees such as {\bf CSS} trees  \cite{rao1999cache}. Answers to those two related questions are not available, although they would be methodologically and practically important. As a matter of fact, and quite surprisingly, those questions have been overlooked so far. 
	
	\subsection{Our  Contributions}
	We settle on the positive both related questions by proposing a novel paradigm in which to cast the design and engineering of Learned Sorted Set  Dictionaries. It is presented in Section  \ref{sec:ModGenDic}, where we also point out that the {\bf RMI} \cite{kraska18case}  and {\bf PGM} \cite{Ferragina:2020pgm}  can be modified to be part of it, those two models being reference in the current State of the Art.
	
	A novel class of models,  which we refer to as {\bf Binning},  that comes out of the paradigm introduced here, is analyzed theoretically in Section 	\ref{sec:fixed}. It naturally lends itself to providing theoretic guarantees in terms of average access time and dynamic operations that are the best known in the Literature. The theoretic part is a natural, although surprisingly overlooked,  generalization of techniques coming from  Dynamic Interpolation Search on non-independent data \cite{Demaine04}, which we cast into a Learned setting. Its methodological merit is to make the theoretic analysis of Learned Sorted Set Dictionaries, an aspect usually addressed poorly, surprisingly simple by allowing the \vir{re-use} of deep results coming from Data Structures. 
	
	As for the experimental part, we consider the static case only, since it is the most consolidated in the Literature, with accepted benchmarks \cite{Kipf19,Mailtry21,Marcus20}.  All our experiments are performed within the Searching on Sorted Sets ({\bf SOSD}  \cite{Kipf19,Marcus20} and {\bf CDFShop}  \cite{Markus20b}  reference software platforms, with the datasets provided there. Data and platforms are available under the GPL-3.0 license. We anticipate that an experimental study concerning the dynamic case is planned. 
	
	Our experiments are reported in Section \ref{sec:exboost} in regard to boosting ability.  We consider the Dictionaries briefly described in Section \ref{Dictionaries}. In regard to them, the {\bf  Binning} model yields consistent boosting. The {\bf PGM} model, via an implementation suitably modified to be part of the paradigm, also provides some boosting, but not for all procedures we have considered. Additional experiments highlight the reasons, bringing to light non-obvious trade-offs between the time to perform a prediction and the time that a procedure takes to search on the full input dataset.

	Finally. in Section  \ref{sec:exspe}, we report experiments clearly showing that the {\bf Binning} Model can yield instances of Learned Sorted Set Dictionaries that are competitive, and many times superior to, benchmark Models such as the  {\bf RMI} and {\bf PGM} in their standard and specific implementations. The source code is available https://github.com/globosco/An-implementation-of-Generic-Learned-Static-Sorted-Sets-Dictionaries).

	\section{From Specific to Generic Learned Dictionaries}\label{sec:GM}
	
	\subsection{Dictionaries Over Sorted Sets}\label{Dictionaries}
	Given an universe $U$ of integers, on which it is defined a total order relation, a \emph{Static Sorted Set Dictionary}
	${\cal SD}$ is a data structure that supports the following operations over a  sorted set  $A \subseteq U$ of $n$ elements: (a) $search(x)=TRUE$ if $x \in A$, otherwise $FALSE$ , (b)  $PSP(x)=\max\{y\in A| y<x\}$, i.e. predecessor search; (c) 
	$range(x,y)= A \cap [x,y]$. The dictionary is  \emph{Dynamic}, if it supports also (d)  $insert(x)$ if $x \notin A$; (e) $delete(x)$ if $x \in A$.

	From now on, for brevity, we refer to ${\cal SD}$  simply as Dictionary, either Static or Dynamic.  Again, for brevity and unless otherwise specified, we consider $search$ only, since $PSP$ is a simple variant of it and $range$ reduces to $search$. 
	
	It is also useful to point out that, for the experimentation regarding this research that,  as already motivated in the Introduction, is limited to Static Dictionaries, we have used the following algorithms and data structures.

	\paragraph{\bf Sorted Table Search Procedures and Variants.} Based on established \vir{textbook material} and accounting for  extensive benchmarking studies present in the Literature \cite{Morin17}, we use the following algorithms. Standard Binary Search \cite{KnuthS}, Uniform Binary Search  \cite{Morin17,KnuthS}, Eytzinger layout Binary Search \cite{Morin17},  B-trees layout Binary Search \cite{Morin17},  and Interpolation Search \cite{Peterson57}. Details regarding their implementation are available in  Section \ref{S-sec:BS} of the Supplementary File. They are abbreviated as {\bf BBS, BFS, BFE, BFT, IS}, respectively.  For {\bf BFT}, a number following the acronym indicates the page size we are using in our experiments. 
	
	%\item [(b)] {\bf Search Trees: Static and Dynamic}.  
	
	\paragraph{\bf Search Trees: Static and Dynamic.} We consider the  CSS Trees  \cite{rao1999cache}, since they are one of the earliest data structures that try to speed up Binary Search, with the use of additional space and being \vir{cache conscious}. We denote them here as {\bf CSS}. As for Balanced Search Trees,  although they have been designed for the dynamic setting of the problems we are considering, they are of interest also in the static case. Among the many possible choices, we consider the Self-Adjusting  Binary  Search Tree \cite{ST85}, denoted {\bf SPLAY}, since it is one of the few data structures that \vir{learns} its  \vir{best} organization from the data. It is to be pointed out that the very well known   {\bf B-Trees} \cite{comer1979ubiquitous} is well represented by the B-trees layout Binary Search.

	\subsection{ Models  Specific For Boosting Binary and Interpolation Search}\label{sec:PS}
	
	Kraska et al. \cite{kraska18case} have proposed an approach that transforms $PSP$  and $search$  into a learning-prediction problem when $A$ is represented as a sorted table. 
	Consider the mapping of elements in the table to their relative position within the table. Since such a function is reminiscent of the  Cumulative Distribution Function over the universe $U$ of elements from which the ones in the table are drawn, as pointed out by Markus et al. \cite{Marcus20}, we refer to it as CDF. Very briefly, the models proposed so far build an approximation of the Cumulative Distribution Function (CDF) of the table, which is used to predict where to search for a given query element.  For the convenience of the reader, a simple example of the learning and query process is provided in Section \ref{S-sec:exquery} of the Supplementary File.  Formal definitions are available in \cite{Marcus20}. 
	The models that have been proposed so far for Learned Sorted Set Dictionaries  (see \cite{amato2021learned,amato2021lncs,Ferragina:2020book,Kipf19,Kipf20,Marcus20}   ), when queried, all return an index  $i$ into the table and an approximation $\epsilon$ accounting for the error in predicting the position of the query element within the table. Then, the search is finalized via Binary Search in the interval $[i-\epsilon, i+\epsilon]$ of $A$.  Other search routines such as Interpolation Search can be used for the final search stage. 
	For this research, and following a recent authoritative benchmarking study \cite{Marcus20}, we take as reference the {\bf RMI}  and the {\bf PGM}  models. For the convenience of the reader, they are described in detail in  Section \ref{S-sec:models} of the Supplementary File, together with an atomic model that summarizes the simple presentation of Learned Dictionaries given so far. 
	
	\subsection{ Models for Boosting Generic Dictionaries}\label{sec:ModGenDic}
	Since we are interested in testing the generality of the \vir{boosting effect} discussed in the Introduction, we introduce a new class of Models. 
	
	\begin{definition}\label{Def:GModels}
		A Generic  Model of type $D$  for sorted set  $A$ is a \vir{black box} that returns an explicit partition of the sorted universe $U$  into intervals, with the elements of $A$  assigned to intervals and kept sorted.
		A visit of the partition from left to right provides $A$. Moreover, in  $O(\log n)$ time, given an element 	$x \in U$, it  
		provides as output the unique interval in the partition where $x$ must be searched for, in order to assess whether it is in $A$ or not. 
	\end{definition}
	
	The main difference between the Models characterized by Definition \ref{Def:GModels} and the ones used so far for Learned Dictionaries is that, apart from the partition of $U$ being explicit,   no two elements can have an intersecting prediction interval. It is to be noted, however, that the {\bf PGM} Index can be easily transformed into a Generic Model. In principle, multi-layer {\bf RMI}s, with a tree structure,  can also be adapted to be Generic Models. However, due to the way they are implemented and \vir{learned} right now,  such a transformation would require a major reorganization of their implementation and \vir{learning} code.  For those reasons, among the Models proposed so far, we consider only the {\bf PGM}, for the experimental part of this research.   
	
	\begin{definition}
		Let  $\hat{D}$ be an instance of a Generic Model of type $D$ for $A$,  consisting of $k$ intervals and let ${\cal SD}$  be a  Dictionary. 
		Its learned version, with respect to model $\hat{D}$ and denoted as $D_k$,  is obtained by building separately an ${\cal SD}$  for each sorted set within each interval in the partition. In order to answer a query,  a prediction query to the model $D_k$    returns a pointer to the predicted interval in the partition and the  ${\cal SD}$   associated with it is queried.
	\end{definition}
	
	Generic  Models can be subdivided into two families: The ones in which the intervals of the partition are of fixed length and the ones in which their lengths are variable. We discuss an example of the first type in Section \ref{sec:fixed}, which is an extension to the Learned setting of a Dynamic Interpolation Search Data Structure proposed by Demaine et al. \cite{Demaine04}.  The method overlooked so far within the development of Learned Sorted Set Dictionaries, is related to the estimation of probability density functions via histograms, e.g, \cite{Freedman1981}.  As for the second type,  and as already anticipated, we consider the {\bf PGM}, suitably modified to fit the new paradigm,  for our experiments. 
	
	\section{Learned Generic  Dictionaries: The Case of Equal Length Intervals via Binning}\label{sec:fixed}
	
	The static case is presented in Sections \ref{sec:CWT}-\ref{sec:EB}, while  Section \ref{sec:Dyn} is dedicated to the dynamic case.  For brevity, the proofs on the Theorems listed next are in Section \ref{S-sec:proofs} of the Supplementary File.

	\subsection{Definition, Construction and Worst Case Search Time}\label{sec:CWT}
	
	Fix an integer $k=O(n)$. 
	The universe  $U$ is divided into $k$ bins $B_1,\cdots, B_k$,  
	each representing a range  of integers of size $\frac{A[n]-A[1]}{k}$.  Each bin has associated the interval of elements of $A$ falling into its range. 
	Let ${\cal SD}$ be a Dictionary. Its Learned version $D_k$ is built as follows. 
	For each of the mentioned intervals, we build a data structure ${\cal SD}$ containing the elements in that interval. Moreover, there is an auxiliary array such that its $j$-th entry provides a pointer to the data structure assigned to bin $j$, which may be empty (no elements in the bin).  
	As for query, given an element $x \in U$, the bin in which we should search is identified via the formula $i= \lceil \frac{(x-A[1])k}{A[n]-A[1]}\rceil$. 
	
	Let  the gap ratio be $\Delta=\frac{G_{max}}{G_{min}}$, where $G$ denotes the distance between two consecutive elements in $A$.  Notice that $ G_{min}>0$, since $A$ is a set, implying that $\Delta$ is finite.  The role of $\Delta$, discussed in Section \ref{sec:delta}, is to \vir{capture} the amount of \vir{pseudo-randomness} in the input data. We have the following. 
	
	\begin{theorem}
		Assume that ${\cal SD}$ can be built in $O(cg(c))$ time on $c$ elements. Assume also that $g(c)$ is  convex and non-decreasing. Then, $D_k$ can be built in $O(ng(n))$ time. Moreover, given an element $x\in U$, the time to search for it in the model $D_k$, built for ${\cal SD}$,  is $O(f( \min (n, \frac{n\Delta}{k}))$, assuming that searching in $\cal{SD}$ can be done $O(f(c))$ time when it contains $c$ elements. 
		\label{th:Dkconstruction}
	\end{theorem}
	
	\subsection{Search Time  and  Input Data Distributions: Smoothness}
	We consider a family of probability distributions, from which the elements in $A$ are drawn from $U$,  that basically \vir{mimics} the nice property of the Uniform Distribution.  Such a family, or specialisations of it, have been used to carry out average case analysis of Static and Dynamic Interpolation Search (see \cite{KAPORIS20}, and references therein). 
	Maximum Load Balls and Bins Chernoff Bounds arguments \cite{Spirakis2008} play a fundamental role here.
	Let $\eta$ be a discrete  probability distribution over the universe $U$, with unkown parametes. Given two function $f_1$ and $f_2$, $\forall x \in  U$, $\eta$ is $(f_1,f_2) $-smooth \cite{KAPORIS20}  if and only if  there exists a constant $\beta$ such  that for all $c_1,c_2,c_3 \in U: c_1 < c_2<c_3$ and for all naturals $\nu \leq n$, for a randomly chosen element ${\cal X}\in U$ it holds: 
	
	\begin{equation}\label{eq:bla}
		{\bf PR}[   c_2 - \lfloor \frac{c_3-c_1}{f_1(\nu)} \rfloor \leq {\cal X} \leq  c_1 | c_1 \leq {\cal X} \leq c_3] \leq \frac{\beta f_2(\nu)}{\nu}
	\end{equation}

	Equation  (\ref{eq:bla}) states that, when we divide the universe $U$ into $f_1(\nu)$ bins, then no bin has probability mass more than  $\frac{\beta f_2(\nu)}{\nu}$. That is, the probability mass is evenly split among the bins.
	
	\begin{theorem}
		
		Assume that the elements  in  $A$ are drawn from  an $(f_1,f_2)$-smooth distribution, with $f_1(n)=cn$, for some constant $ 0< c \leq 1$,   and $f_2(n)= O( \ln^{O(1)} n)$. Then, 
		given an element $x\in U$,  the time to search for it in the Learned Dictionary $D_{f_1(n)}$, built for ${\cal SD}$,  is $O(f(\min (\log^{O(1)} n, \frac{\Delta}{c}))$,  with high probability (i.e. $1- o(1)$) and assuming that searching in $\cal{SD}$ can be done in $O(f(m))$ time when it contains $m$ elements. 
		\label{th:smoothsearch}	
	\end{theorem}
	
	It is to be noted that the Uniform Distribution is a smooth distribution according to the definition given above. Other smooth distributions of interest are mentioned in \cite{KAPORIS20}. For those distributions,  the Binning-based Learned Dictionary $D_k$, with $k=cn$ and $ 0<c \leq  1$,  can be queried in $O(\ log \ log n)$ expected time via a terminal stage of Binary Search, generalizing well-known results regarding Interpolation Search, e.g., \cite{Gonnet80,Perl78,YY76}.  
	
	\subsection{Search Time and Input Data Distributions: Non-Independence,  Real Datasets and the Role of $\Delta$}
	\label{sec:delta}
	
	In the original proposal by Demaine et al. \cite{Demaine04} of their bin data structure that we have extended here, the role of $\Delta$  is meant to allow an extension of the good behaviour of Interpolation Search on input data drawn uniformly and at random from $U$ to non-random ones, in particular non-independent.  That is, $\Delta$ is supposed to capture the amount of  \vir{pseudo-randomness} in the data. To this end, 
	Demaine et al. showed that, when the input data is drawn from the Uniform Distribution and for $k=n$, then  $\Delta=O(poli\log n)$.  That is, if we consider only $\Delta$ in the complexity of searching in  $D_n$,  for Uniform input data Distributions, then the well known  $O(\log \log n)$ time bound for Interpolation Search would hold for $D_n$. 
	Unfortunately, their theoretic result is non easily extendable, if at all, to Probability Distributions other than Uniform. Therefore, we study $\Delta$ experimentally and on real benchmark datasets. The experiment we have performed, due to space limitation, is described with its outcome, in Section \ref{S-sec:experiments} of the Supplementary File.  We limit ourselves to report that the $poli \log n$ behaviour of  $\Delta$ depends on the dataset, rather than on the CDF and PDF  characterizing the dataset.  However, there is no harm in using it. Those experiments hint at the fact that $\Delta$ may play some role, depending on the specific dataset. To the best of our knowledge, we provide the first study of this important parameter. 
	
	\subsection{Search Time  and  Query Distributions: Learned Optimal Binary Search Forest   with Entropy Bounds}\label{sec:EB}
	
	Assume now that we are given $2n+1$ probabilities, $p_1,p_2, \ldots, p_n$ and $q_0, q_1, \ldots, q_n$, where each $p_i$ is the probability of searching for $A[i]$, $q_0$ is the probability of searching for an element less than $A[1]$, $q_n$ the probability of searching for an element larger than $A[n]$, and $q_i$ is  the probability of searching for an element between $A[i]$ and $A[i+1]$. We assume that the sum of all those probabilities is one. Moreover, we denote the first part of the distribution by $P$ and the second one by $Q$.  Recall from the Literature \cite{Knuth71,NAGARAJ97,Yao80}, that the cost of a Binary Search Tree $T$ is is defined as $C(T)=\sum_{i=1}^{n}p_i\times (depth(A[i])+1)) +\sum_{i=0}^{n}q_i\times depth(i)$, where $depth(A[i])$ is the depth of the node storing $A[i]$ in $T$ and $depth(i)$ is the the depth of the external leaf (finctuous) corresponding to an unsuccessful  search. 
	The Optimum Binary Search Tree, or good approximations of it,  can be found efficiently  \cite{NAGARAJ97}.  Assuming for simplicity that no bin is empty, we extend the notion of Optimum Binary Search Tree to Learned Optimum Binary Search Forest as follows. 
	
	Let $\ell_1,\ell_2, \ldots, \ell_k$ be the number of elements in the bins.
	Consider bin $B_s$ and assume that it contains all elements in $A[i,j]$. The weight of $B_s$ can be defined as $W(B_s)= P_{i,j}+ Q'_{i,j}$ where $P_{i,j}$ is the portion of $P$ restricted to the elements in the bin.
	As for
	$Q'$, it is as $Q_{i,j}$ (the analogous of $P_{i,j}$), except that the failure probabilities preceding $p_i$ and succeeding $p_j$ are lower than the corresponding ones in $Q$: elements not present may fall in part in a bin and in part in its neighbours.  Note that $W(B_i)$ can be interpreted as the probability of searching for an element in $B_s$. 
	A Learned Binary Search Forest can readily be obtained by storing the elements of each bin into a Binary Search Tree.
	Its cost can be defined as $C(D_k,T)= \Sigma_{i=1}^k W(B_i)(1+C(T(B_i)))$, where $T(B_i)$ is the cost of  a Binary Search Tree  storing the elements in $B_i$, with access weigths $P_{i,j}$ and $Q'_{i,j}$.  Let $H(P,Q)$ be the entropy of the probability distribution given by $P$ and $Q$. 
	
	\begin{theorem}
		Consider the access probabilities $P$ and $Q$, defined earlier,  to elements of $A$ and let the interval numbers in a partition be bounded by $k_{\max}$. We have (a) for $ 1 \leq  k \leq k_{\max}$,  the Optimum  Learned Binary Search Forest, i.e, the one with the best cost among all $k$'s,   can be computed in $O(n^2k_{max})$ time, and its cost is bounded by $H(P,Q)+2$; (b) an approximate Optimum  Learned Binary Search Forest, with cost bounded as in the optimum case, in $O(nk_{\max})$ time. 
		\label{th:entropybounds}
	\end{theorem}
	
	It is to be noted that, to the best of our knowledge, the {\bf PGM} model is the only one for which one can prove \vir{entropy bounds} on access costs, but only for successful searches. The analysis reported in \cite{Ferragina:2020pgm} is not as general as the one reported here in terms of success and failure search probabilities and optimality is not considered. 
	
	\subsection{The Dynamic Case}\label{sec:Dyn}
	Again, this is a variant of the Dynamic Case of Interpolation Search proposed by Demaine et al. in Section  3 of their paper \cite{Demaine04}.  Letting  ${\cal SD}$ now denote the dynamic dictionary to be coupled with $D_k$. which is built as in the static case, with a few important changes. Elements in each bin are allocated to a copy of  ${\cal SD}$ each. 
	Let $\hat{ \Delta}$ be the  gap ratio computed when the structure is initially built. During its evolution, the structure will be valid for values of $\Delta \leq  \hat{ \Delta}$. When this condition no longer holds, $\hat{\Delta}$ is recomputed from the data in the structure, as specified shortly. The range being considered is now $[A[1]- L\hat{\Delta}, A[n]+L\hat{\Delta}]$, where $L=A[n]-A[1]$. That  inteval is subdivided into $k$ bins, each of size 
	$\frac{(2\hat{\Delta}+1))L}{k}$. The structure is rebilt when $n/2$ updates have taken place without a rebuild or the value of  $\hat{\Delta}$ is no longer valid, i.e., the gap ratio has gone up due to the updates. In this latter case, we rebuild  the data structure  with $\hat{\Delta}= \max(\Delta_{new}, 2{\hat \Delta_{old}})$, where $\Delta_{new}$ is computed from the data present in the structure and $\hat \Delta_{old}$ is the gap ratio of the last rebuild.
	
	\begin{theorem}\label{th:dynamic}
		Assume that ${\cal SD}$ can be built in linear time and that it supports  $search$, $insert$ and $delete$ operations in  $\log$ time.  	For the Learned Dictionary $D_k$, the following bounds hold. The worst case cost of $search$, $insert$ and $delete$
		is $O(\log \min(n, \frac{n}{k}\Delta_{max})$, where $\Delta_{max}$ is the maximum $\Delta$ achieved over the lifetime of the structure.  An analogous bound applies to the case in which the operations can be done in $\log \log$ time. The  rebuilding of the data structure takes an amortized cost per element of $O(\log \frac{n}{k}\Delta_{max})$ time. When $\Delta_{max}$ is known \emph{a priori},  the total cost of rebuild is now linear in the number of elements in the data structure, over its lifetime.
	\end{theorem}
	
	\begin{corollary}\label{co:boh2}
		When  $|U|=O(n^c)$, $c$ constant, under the same assumpions of Theorem \ref{th:dynamic}, $search$, $insert$, $delete$ have the same worst case time bounds. 
		The rebuilding of the data structure takes an amortized cost per element of $O(\log n)$ time. 
	\end{corollary}
	
	It is to be noted that Corollary \ref{co:boh2} provides the first results concerning Dynamic Learned Sorted Set Dictionaries with logarithmic performance, both in terms of operations and rebuild. This is analogous to what one gets with classic data structures and the result here is methodologically important. Prior to this, the {\bf PGM} Model is the only one for which provably good bounds have been given, but $search$ is not $O(\log n)$ worst-case time, as here. Finally, the assumption that the size of the Universe is polynomially related to the size of the dataset is a widespread and commonly accepted assumption, that has its roots both in \vir{theory} and \vir{practice}, e.g., see discussion in \cite{KAPORIS20}. 
	
	\section{Experimental Methodology}\label{sec:expm}
	All the experiments have been performed on a workstation equipped with an Intel Core i7-8700 3.2GHz CPU with  32 Gbyte of DDR4. The operating system is Ubuntu LTS 20.04.
	We use the same real datasets of the benchmarking study on Learned Indexes \cite{Kipf19,Marcus20}. In particular, we restrict attention to integers only, each represented with 64 bits, since the datasets with 32 bits add no discussion to the experiments.  
	For the convenience of the reader, a list of those datasets, with an outline of their content, is provided next. They are: (a) {\bf amzn}: book popularity data from Amazon (each key represents the popularity of a particular book); (b)  {\bf face}: randomly sampled Facebook user IDs, (each key uniquely identifies a user); (c)   {\bf osm}: cell IDs from Open Street Map  ( each key represents an embedded location); {\bf	wiki}: timestamps of edits from Wikipedia  (each key represents the time an edit was committed). Each dataset consists of 200 million elements for roughly 1.6Gbytes occupancy.  As for query dataset generation, for each of the tables mentioned earlier, we extract uniformly and at random (with replacement) from the Universe $U$ a total of two million elements, 50\% of which are present and  50\% absent,  in each table. All these datasets are available in https://osf.io/ygnw8/?view\_only=f531d074c25b4d3c92df6aec97558b39. 
	
	\section{Experiments: Results and Discussion}

	\subsection{Boosting}\label{sec:exboost}
	
	As anticipated in Section \ref{sec:GM}, we investigate whether Generic Learned Dictionaries provide the \vir{boosting effect} already known in the Literature for Specific Learned Dictionaries. So, as anticipated in Section \ref{sec:ModGenDic},  we analyze the following two cases, in the static scenario.
	
\begin{figure}[tbh]
		%\begin{center}
		\centering
		(a)
		\begin{minipage}{0.45\textwidth}
			\includegraphics[width=\linewidth]{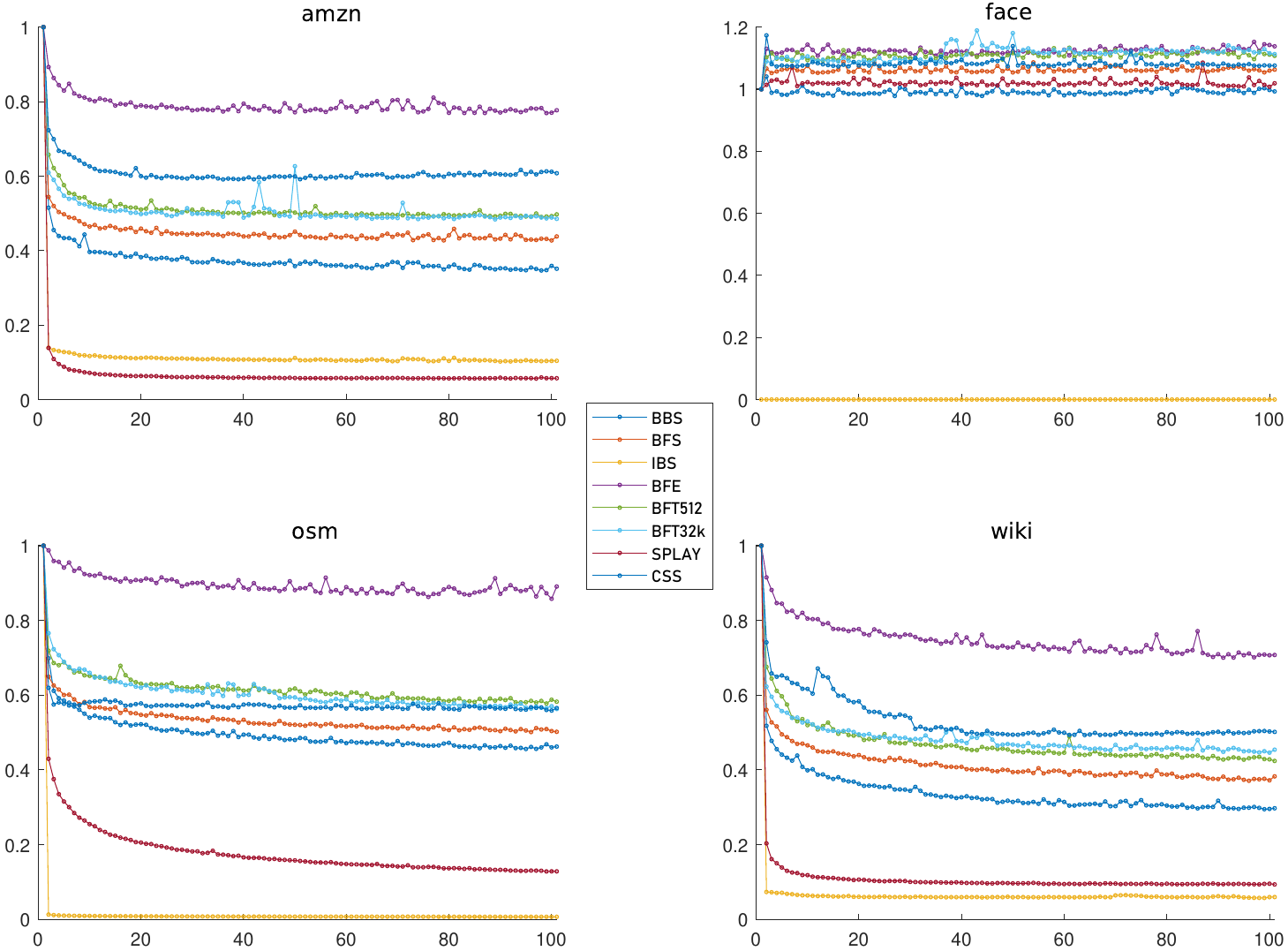}
		\end{minipage}\hfill
		(b)
		\begin{minipage}{0.45\textwidth}
			\includegraphics[width=\linewidth]{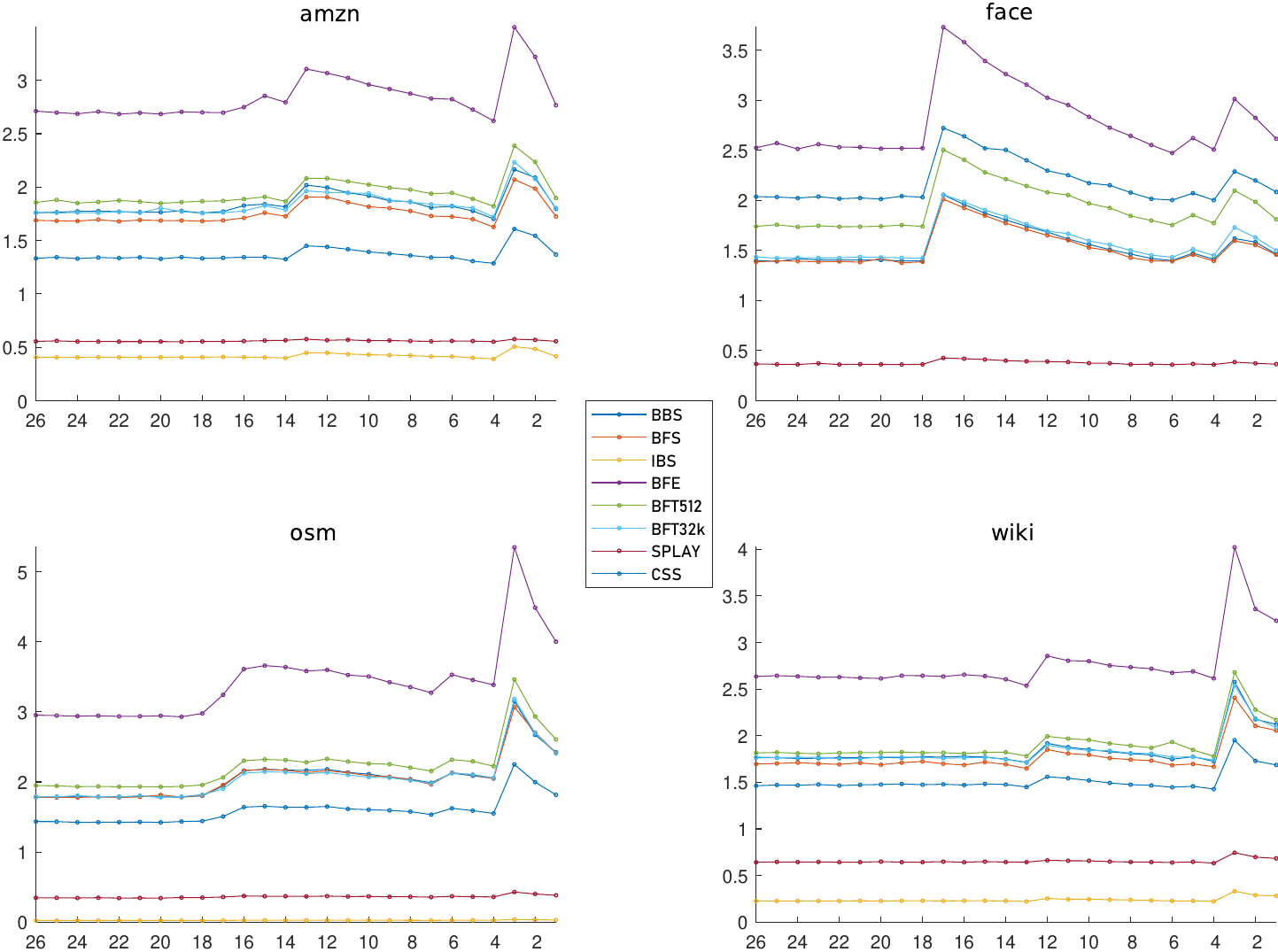}
		\end{minipage}\hfill
		\caption{{\bf Boosting Property for Binning and PGM Models}. The Dictionaries considered for  those experiments are provided in the Legend. (a) The $x$ axis reports, for each dataset, the number of bins in percentage with respect to the number of elements in the sorted set. Given a dictionary ${\cal SD}$, the $y$ axis shows the ratio between the mean query time of the {\bf Binning} Learned version $D_k$ and $\cal{SD}$ alone. A $y$ value below one indicates the superiority of the {\bf  Binning } Learned Model  vs ${\cal SD}$ alone. (b) The $x$ axis reports, for each dataset, the chosen $\epsilon$ for the {\bf PGM} construction. The $y$ axis  is  analogous to the {\bf Binning } case.  }
		%\end{center}
		\label{fig:boosting}
	\end{figure}

	\paragraph{{\bf Binning.}} Fix a Dictionary ${\cal SD}$, chosen from the ones used for this research and described in Section \ref{Dictionaries} and consider its Learned Generic version obtained via {\bf  Binning}.  For each dataset, we increase the space occupied by the Generic Learned version, growing the number of bins in percentage with respect to the number of elements in the given Tables, i.e. from 0\% to 100\%, and we calculate the ratio between the mean query time of a Generic Learned version $D_k$  and its non-learned version ${\cal SD}$. A ratio below one indicated that the Generic Learned version boosts the performance of the non-learned version.  The results are reported in Figure \ref{fig:boosting}(a). As it is evident from that Figure, all the Data Structures chosen for this research register a boosting effect, except for the  {\bf face} dataset. The explanation is quite simple. Although the CDF of {\bf face}, shown in Figure  \ref{S-fig:cdf_plots} of the Supplementary File,  provides the impression of \vir{uniformity}, there are a few outliers that cause most of the bins to be empty, with most of the elements falling in a few bins. That is, the {\bf  Binning} method is sensible to outliers that unbalance the binning. Fortunately, for other real CDFs (see again Figure  \ref{S-fig:cdf_plots} in the Supplementary File), even as complicated as the one of {\bf osm}, the improvement is quite relevant. With reference to those datasets, it is quite impressive that the improvement follows the same shape in all of them. Moreover, most of the gain of using a {\bf  Binning}  Learned Dictionary over the standard Dictionary is concentrated around a low percentage of bins being used. That is, \vir{the spreading of elements over the bins} has a \vir{diminishing return}in terms of mean query time,  as the number of bins grows.
	
	\paragraph{{\bf PGM}.}  The division of the Universe  $U$ into intervals now depends on the approximation parameter $\epsilon$ (see the description of this Model in Section {\ref{S-sec:models} of the Supplementary File}).  We proceed as follows. For each dataset, we choose $\epsilon$ as a power of two in the interval $[1, \frac{n}{2} ]$. That is, we built models with an increasing error that partitions the Universe $U$ from very small intervals to the one that contains the whole dataset. Then, we proceed as in the {\bf  Binning} case, given a dictionary ${\cal SD}$.  The results are reported in Figure \ref{fig:boosting}(b). As evident from that Figure, we have a boosting effect only for {\bf IBS} and {\bf SPLAY}. In order to gain insights into the reason for that, we have performed additional experiments, involving {\bf Binning} and the {\bf PGM}. They are reported in Figures \ref{S-fig:bin_time}-\ref{S-fig:pgm_time_splay} of the Supplementary File. For the {\bf Binning} Model,  the meantime to perform a prediction is negligible with respect to the subsequent search stage on a reduced set of data, i.e., the search routine can take full advantage of the reduction in the size of the dataset to search into. It is worth recalling that the prediction in this model takes $O(1)$ time. As for the {\bf PGM}, a prediction can be made by \vir{navigating} a tree (see the description of the {\bf PGM} in Section  \ref{S-sec:models} of the Supplementary File), i.e., it is no longer a constant since it depends on the number of segments in which $U$ has been divided (see the analysis in 
	\cite{Ferragina:2020pgm}). Although the reduction in dataset size may be substantial (data not shown and available upon request), it is now the trade-off between prediction time and search in the full set that determines the boosting effect. 
	%\end{enumerate}
	
	In conclusion, those experiments hint at the fact that we get a boosting effect out of Generic Learned Models, as long as the \vir{navigation} time to get to the appropriate interval is rather small, compared to a full-fledged search into the original dataset. It is to be noted that a two-level  {\bf RMI}, which is the one recommended for use in \cite{Marcus20},  also has the potential for an $O(1)$ \vir{navigation}. However, due to its current highly engineered implementation,   is not so immediate to transform that software into a Generic {\bf RMI }Model, without risking to compromise its query efficiency.  
	
	\subsection{Competitiveness of Generic Learned Dictionaries with Respect to Specific Ones}\label{sec:exspe}
	
	These experiments provide a comparison with the Models that well represent the State of the Art, i.e. {\bf RMI} and {\bf PGM}.  In particular, in the following, we discuss two possible scenarios. 
	
\begin{figure}[tbh]
		%\begin{center}
		\centering
		(a)
		\begin{minipage}{0.25\textwidth}
			\includegraphics[width=\linewidth]{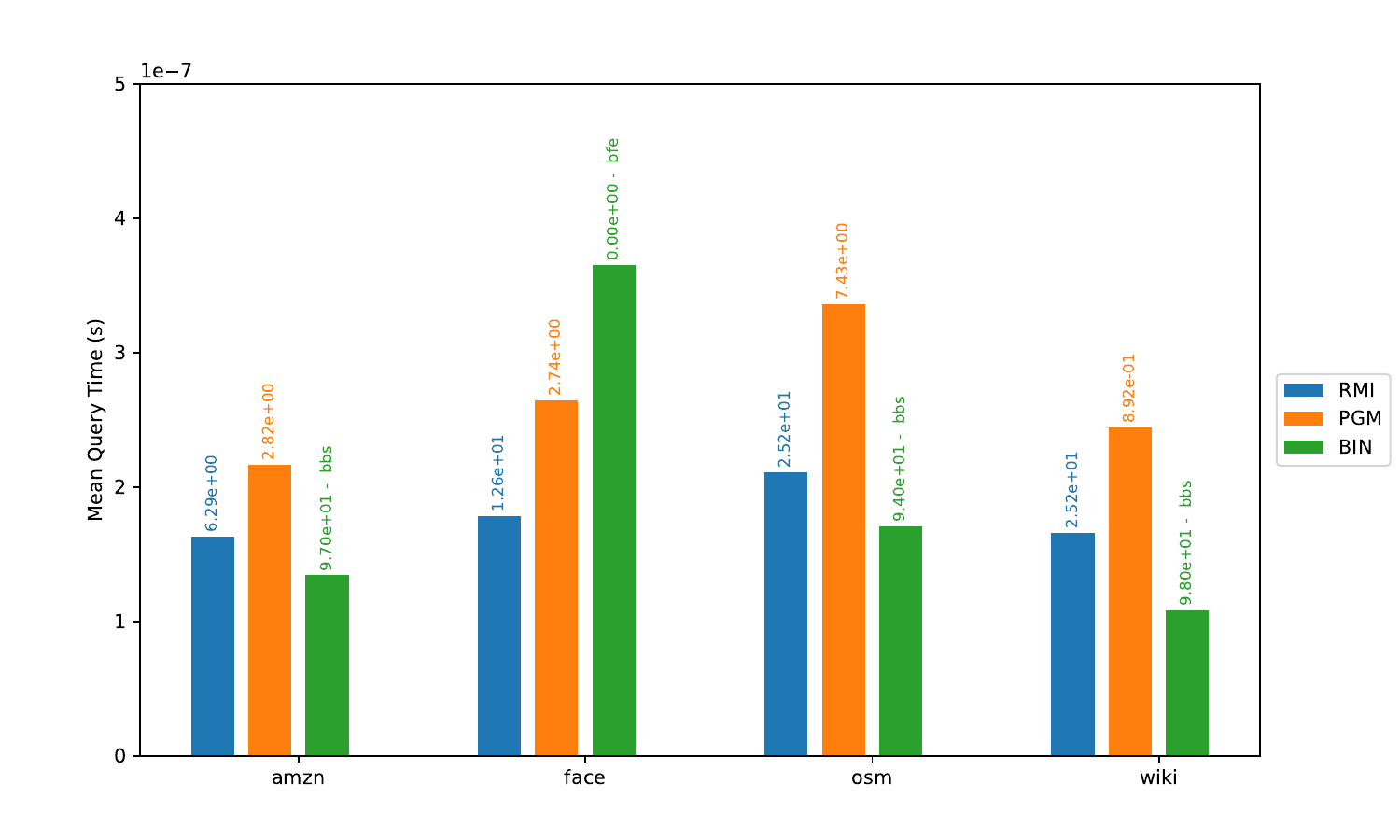}
		\end{minipage}\hfill
		(b)
		\begin{minipage}{0.6\textwidth}
			\includegraphics[width=\linewidth]{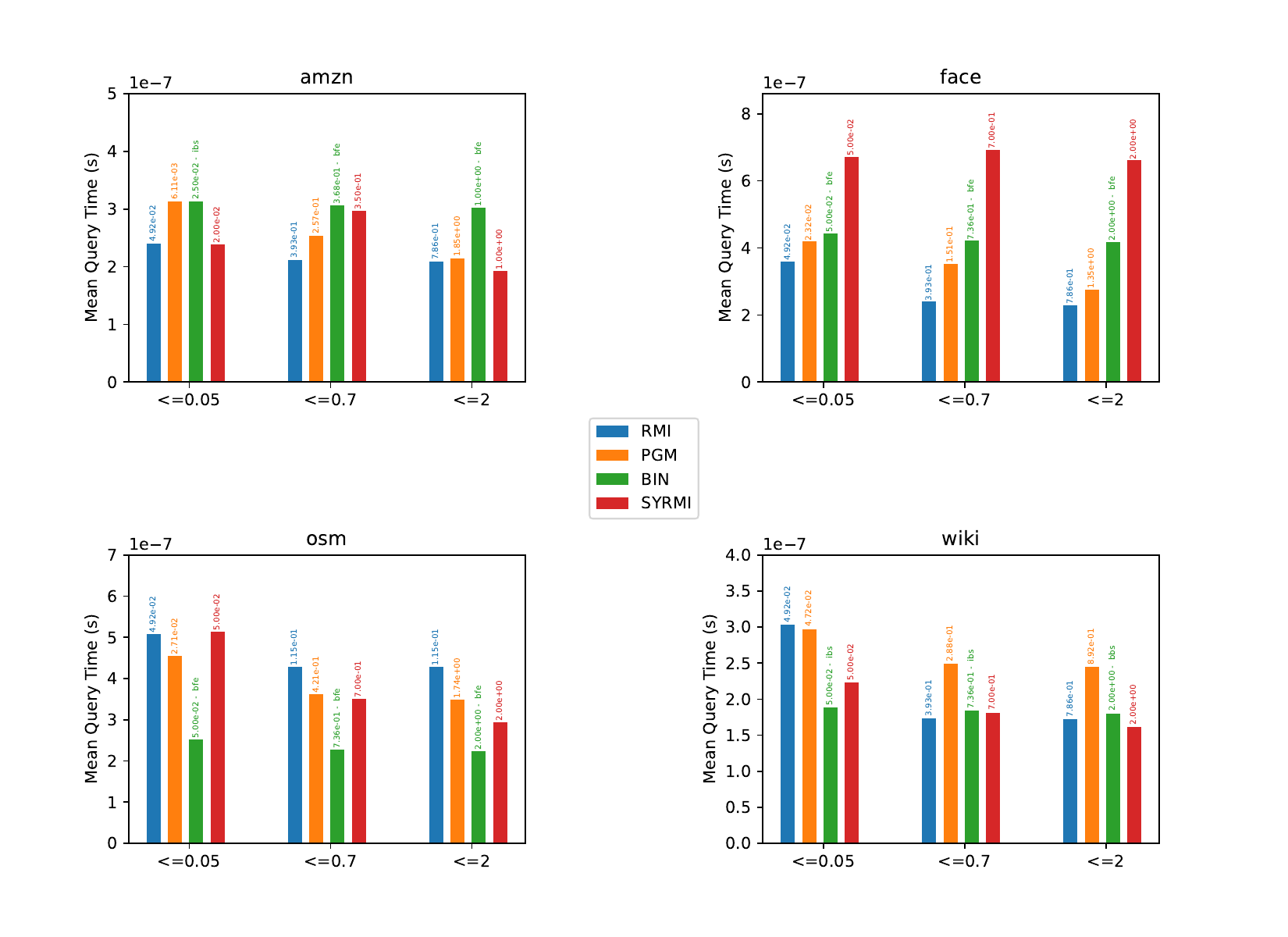}
		\end{minipage}\hfill
		
		\caption{{\bf Learned Indexes Query Time.} (a) No Bounds on Space. For each dataset, we report the mean query time of the best Learned Indexes including the Generic Learned Dictionary denoted with {\bf  BIN}. Above each bar, we report the model space used as a percentage of the table size. For  {\bf  BIN}, it is also indicated which dictionary performs best. (b) With Bounds on Space. For each dataset, we choose three space bounds as detailed in the main text.  For each space-bound, from left to right, we report the mean query time of the best {\bf  RMI} and the {\bf  PGM}  that satisfies the imposed bound. The next bar is the mean query time for the best {\bf  Binning}  Generic Learned Dictionary that satisfies the given bound on space.  The last bar indicates the mean query time for the {\bf  SY-RMI} that satisfies the bound on space.}
		%\end{center}
		\label{fig:CompAll}
	\end{figure}

	\paragraph{ No Bounds on Model Space.} In this case, how much space the model uses with respect to the one occupied by the input sorted set is not a critical requirement, i.e., query time is privileged.   For each dataset, we have built a {\bf Binning} Learned Dictionary for each  Dictionary ${\cal SD}$ used for the experiments in this research. Then, among all these Learned Dictionaries, we have chosen the one with the smallest mean query time. For each dataset, we have also trained  {\bf RMI} and {\bf PGM} models, in agreement with the benchmark  procedures adopted in  \cite{amato2021learned,amato2021lncs,Marcus20}, i.e., we have taken the top (at most) ten performing models considered by {\bf SOSD}. We have then chosen, for each of the considered models, the one that provides the best mean query time. The  results are reported in Figure  \ref{fig:CompAll}(a).  Given the fact that the  {\bf  RMI} and {\bf  PGM}  are among the best performing models in the Literature, it is quite remarkable that the {\bf Binning } strategy, with {\bf  BBS} as the final search routine, outperforms those models on three of the four benchmark datasets. Its relatively poor performance on the {\bf  face} datasets with respect to the {\bf RMI} is motivated by the presence of outliers that, as already observed, prevent to take full advantage of the {\bf Binning } strategy. Another notable fact is that the use of layouts other than sorted, e.g., {\bf BFE} may be of help. Those layouts cannot be used within 
	the current Specific Models. Interestingly, with reference to the CDF figures reported in Fig. \ref{S-fig:cdf_plots} of the Supplementary File, the {\bf Binning} strategy is able to perform well on CDFs of different nature. On the \vir{munus} side, the {\bf Binning } approach tends to use more space than the other best models.

	\paragraph{\bf   Bounds on Model Space.} A model that guarantees a good mean query time in small additional space, with respect to the one taken by the input sorted set,  is desirable in many applications \cite{amato2021learned,amato2021lncs,Ferragina:2020pgm}.  Therefore, for each dataset,  we now impose three space bounds that a model must satisfy, expressed in percentage with respect to the size that the sorted set would occupy if stored in an array. Namely, $0.05, 0.07, 0.2\%$.    Then, for each percentage, we have selected the best mean query time {\bf  Binning} Learned Dictionary that satisfies the given percentage bound. As for the other Models considered in this research, we have done the same selection, reporting the performance of the best model, for each space bound. Finally, we have also considered the {\bf SY-RMI} Model, which belongs to the class of {\bf  RMI} Models, and that has been specifically designed to yield good mean query times in small space  \cite{amato2021learned,amato2021lncs}.  The results are reported in Figure \ref{fig:CompAll}(b). Those results bring to light that, even in small space, the {\bf  Binning } strategy is competitive with respect to reference models, in particular when used on datasets with a complex CDF, e.g., {\bf osm}. Again, it is of interest to notice that competitive performances are obtained via the use of {\bf BFE} for the final search stage, a routine that can be used by current Specific Models. 
	
	\section{Conclusions and Future Directions}
	We have provided a new paradigm for the design of Learned Dictionaries. As opposed to the current state of the Art, it can be applied to any Sorted Set Dictionary, rather than to only search routines with a sorted layout. The theoretic analysis performed for the {\bf Binning} Model shows that we can leverage on classic results from Data Structures to obtain sound evaluations of the performance of Learned Dictionaries, an aspect usually addressed poorly. We have also given experimental evidence that the new paradigm, as far as the static case is concerned, can yield valid Data Structural  Boosters and be competitive with reference solutions available in the Literature.
	For the future, the dynamic case is to be considered. It implies a careful re-design of the software solutions available so far, i.e., 
	\cite{Ding20,Ferragina:2020pgm} as well as a well planned experimental setting since it is not clear that the one available for the static case may be the best choice to study the dynamic case. 
	We point out that the societal impacts of our contribution are in line with general-purpose Machine Learning technology.

	\bibliographystyle{plain}
	\bibliography{references}

\end{document}

\title{From Specific to Generic Learned Sorted Set Dictionaries: A Theoretically Sound Paradigm Yelding Competitive Data  Structural Boosters in Practice -Supplementary  File}
	
\author{Domenico Amato\\
	\and
	Giosu\'e Lo Bosco\\
	\and
	Raffaele Giancarlo}

\date{
	Dipartimento di Matematica e Informatica\\ 
	Universit\'a degli Studi di Palermo, ITALY\\
	%{\{domenico.amato,andrea.desalve,giosue.lobosco,raffaele.giancarlo\}@unipa.it}\\
	\today
}

	\maketitle
	
	\begin{abstract}
		This document provides additional details with respect to the content of the Main Manuscript by the same title.
	\end{abstract}

	\section{ Sorted Table Search Procedure and Variants: Details}\label{sec:tableproc}

	\subsection{ Array Layouts for Binary Search}\label{sec:BS}
	
	Following research in \cite{Morin17}, and with the use of the shorthand notation in the Main Text (Section \ref{M-Dictionaries}), we review here three basic layouts.
	
\begin{figure}[]
		\begin{center}
			\includegraphics[width=0.40\textwidth]{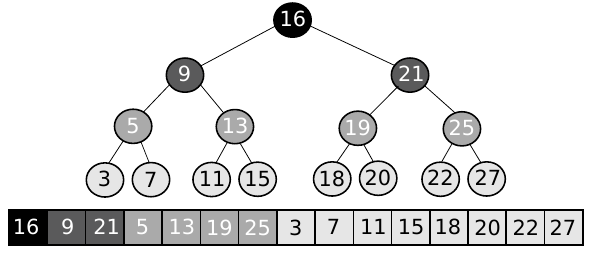}
			\caption{{\bf An example of  Eyzinger layout of a table with 15 elements}. See also \cite{Morin17}.}
			\label{fig:EyL}
		\end{center}
	\end{figure}

\begin{figure}[tbh]
		\begin{center}
			\includegraphics[width=0.40\textwidth]{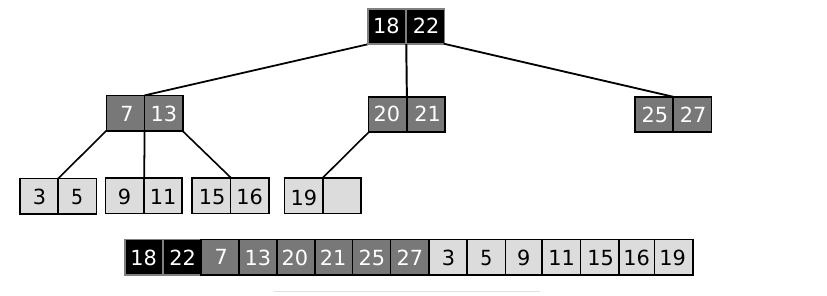}
			\caption{An example of $(B+1)$ layout of a table  with 15 elements and $B=2$ (see also \cite{Morin17}).}
			\label{fig:B-Tree}
		\end{center}
	\end{figure}

	\begin{itemize}
		
		\item [1] Sorted. It is the classic textbook layout for standard Binary Search,  reported in Algorithm  \ref{AL:BBS}. We also consider an implementation of Uniform Binary Search \cite{Morin17,KnuthS}) 
		(Algorithm \ref{AL:BFS}), that uses prefetching. This routine is referred to as branch-free in \cite{Morin17}, since there is no branching in the while loop. Indeed, as explained in \cite{Morin17}, the case statement in line 8 of Algorithm \ref{AL:BFS} is translated into a conditional move instruction that does not alter the flow of the assembly program corresponding to that C++ code. In turn, that has the advantage of a better use of instruction pipelining with respect to the branchy version of the same code. Prefetching, i.e., instructions 6 and 7 in Algorithm  \ref{AL:BFS}, refers to the fact that the procedure loads in the cache data ahead of their possible use. As shown in \cite{Morin17}, such an implementation of Binary Search is substantially faster than its branchy counterpart.

		\item [2] Eytzinger Layout \cite{Morin17}.  The sorted table is now seen as stored in a virtual complete balanced binary search tree. Such a tree is laid out in Breadth-First Search order in an array. An example is provided in Fig. \ref{fig:EyL}. Also, in this case, we adopt a branch-free version with prefetching of the Binary Search procedure corresponding to this layout. It is reported in Algorithm \ref{AL:BFE}.

		\item [3] B-tree. The sorted table is now seen as a $B+1$ search tree \cite{comer1979ubiquitous}, which is then laid out in analogy with an Eytzinger layout.  An example is provided in Fig. \ref{fig:B-Tree}. Also, in this case, we adopt a branch-free version of the Binary Search procedure corresponding to this layout, with prefetching. It is taken from the software associated with \cite{Morin17} and it is reported Algorithm \ref{AL:BFT}.

	\end{itemize}
	
	\begin{algorithm}
		\caption{{\bf Implementation of Standard Binary Search}.}\label{AL:BBS}
		\begin{algorithmic}[1]
			\BState int StandardBinarySearch(int *A, int x,  int left, int right)\{
			\State  \ \ \ while (left $<$ right) \{
			\State	\ \ \ \ \ int m = (left + right) / 2
			\State	\ \ \ \ \ if(x $<$ A[m])\{
			\State  \ \ \ \ \ \ \ rigth = m;
			\State  \ \ \ \ \ \}else if( x $>$ A[m])\{
			\State  \ \ \ \ \ \ \ left = m+1;
			\State  \ \ \ \ \ \}else\{
			\State  \ \ \ \ \ \ \ return m
			\State  \ \ \ \ \ \}
			\State \ \ \ \}
			\State \ \ \ return right;
			\BState \}	
		\end{algorithmic}
	\end{algorithm}
	
	\begin{algorithm}
		\caption{{\bf Implementation of Uniform Binary Search with Prefetching}. The code is as in   \cite{Morin17} (see also \cite{KnuthS})}.
		\label{AL:BFS}
		\begin{algorithmic}[1]
			\BState int prefetchUniformBinarySearch(int *A, int x,  int left, int right)\{
			\State \ \ \ const int *base = A;
			\State \ \ \ int n = right;
			\State \ \ \ while (n $>$ 1) \{
			\State	\ \ \ \ \ const int half = n / 2;
			\State	\ \ \ \ \ \_\_builtin\_prefetch(base + half/2, 0, 0);
			\State	\ \ \ \ \ \_\_builtin\_prefetch(base + half + half/2, 0, 0);
			\State	\ \ \ \ \ base = (base[half] $<$ x) ? \&base[half] : base;
			\State	\ \ \ \ \ n -= half;
			\State	\ \ \ \}
			\State \ \ \ return (*base $<$ x) + base - A;
			\BState \}	
		\end{algorithmic}
	\end{algorithm}
	
	\begin{algorithm}
		
		\caption{{\bf  Implementation of Uniform Binary Search with Eytzinger Layout and Prefetching}. The code is as in   \cite{Morin17}.}
		\label{AL:BFE}
		\begin{algorithmic}[3]
			\BState int UniformEytzingerLayout(int *A, int x,  int left, int right)\{
			\State \ \ \ int i = 0;
			\State \ \ \ int n = right;
			\State \ \ \ while (i $<$ n)\{
			\State \ \ \ \ \ \_\_builtin\_prefetch(A+(multiplier*i + offset)); 
			\State \ \ \ \ \ i = (x $<=$ A[i]) ? (2*i + 1) : (2*i + 2);
			\State \ \ \ \}
			\State \ \ \ int j = (i+1) $>>$ \_\_builtin\_ffs($\sim$(i+1));
			\State \ \ \ return (j == 0) ? n : j-1;
			\BState \}	
		\end{algorithmic}
	\end{algorithm}
	
	\begin{algorithm}
		
		\caption{{\bf Implementation of Uniform Binary Search with B-Tree Layout and Prefetching}.} \label{AL:BFT}
		\begin{algorithmic}[1]
			\BState int UniformBTreeSearch(int* a, int x,  int left, int right)\{
			\State \ \ int j = right;
			\State \ \ int i = left;
			\State \ \ int n = right - left +1;
			\State \ \ int B = cacheline/sizeof(T);
			\State \ \ while (i + B $\leq$ n) \{
			\State \ \ \ \ \_\_builtin\_prefetch(a+child(i, B/2, B), 0, 0);
			\State \ \ \ \ const int *base = \&a[i];
			\State \ \ \ \ const int *pred = uniform\_inner\_search(base, x, B);
			\State \ \ \ \ int nth = (*pred $<$ x) + pred - base;
			\State \ \ \ \ \{
			\State \ \ \ \ \ \ const int current = base[nth \% B];
			\State \ \ \ \ \ \ int next = i + nth;
			\State \ \ \ \ \ \ j = (current $\geq$ x) ? next : j;
			\State \ \ \ \ \}
			\State \ \ \ \ i = child(nth, i, B);
			\State \ \ \}
			\State \ \ if (\_\_builtin\_expect(i $<$ n, 0)) \{
			\State \ \ \ \ 	const int *base = \&a[i];
			\State \ \ \ \ 	int m = n - i;
			\State \ \ \ \ 	while (m $>$ 1) \{
			\State \ \ \ \ \ \ int half = m / 2;
			\State \ \ \ \ \ \ const int *current = \&base[half];		
			\State \ \ \ \ \ \ base = (*current $<$ x) ? current : base;
			\State \ \ \ \ \ \ m -= half;
			\State \ \ \ \ \}
			\State \ \ \ \ 	int ret = (*base $<$ x) + base - a;
			\State \ \ \ \ 	return (ret == n) ? j : ret;
			\State \ \ \}
			\State \ \ return j;
			\BState \}
			\State
			\BState int* uniform\_inner\_search(int *base, int x, int C) \{
			\State \ \	if (C $\leq$ 1) return base;
			\State \ \	const int half = C / 2;
			\State \ \	const T *current = \&base[half];
			\State \ \	return uniform\_inner\_search((*current $<$ x) ? current : base, x, C-half);
			\BState	\}
		\end{algorithmic}
	\end{algorithm}
	
	\subsection{Interpolation Search}\label{sec:INT}
	
	As stated in the Main Text, such a Sorted Table Search technique was introduced by Peterson \cite{Peterson57} and it has received some attention in terms of analysis \cite{dataStructHand}, since it works very well on data following a Uniform or nearly Uniform distribution and very poorly on data following other distributions. The procedure adopted here is reported in Algorithm \ref{AL:IBS}. 
	
	\begin{algorithm}
		\caption{{\bf Implementation of Standard Interpolation Search}.}\label{AL:IBS}
		\begin{algorithmic}[1]
			\BState int StandardInterpolationSearch(int *arr, int x, int start, int end)\{
			\State \ \ \ int lo = start, hi = (end - 1);
			\State \ \ \ while (lo $\leq$ hi \&\& x $\geq$ arr[lo] \&\& x $\leq$ arr[hi]) \{
			\State	\ \ \ \ \  if (lo == hi) \{
			\State	\ \ \ \ \ \ \ if (arr[lo] == x) return lo;
			\State	\ \ \ \ \ \ \ return -1;
			\State	\ \ \ \ \ \}
			\State	\ \ \ \ \ int pos = lo + (((double)(hi - lo) / (arr[hi] - arr[lo])) * (x - arr[lo]));
			\State	\ \ \ \ \ if (arr[pos] == x) return pos;
			\State	\ \ \ \ \ if (arr[pos] $<$ x) lo = pos + 1;
			\State	\ \ \ \ \ \ \ else hi = pos - 1;
			\State	\ \ \ \}
			\State \ \ \ return pos;
			\BState \}	
		\end{algorithmic}
	\end{algorithm}
	
	\begin{figure}[tbh]
		%\begin{center}
		\centering
		(a)
		\begin{minipage}{0.25\textwidth}
			\includegraphics[width=\linewidth]{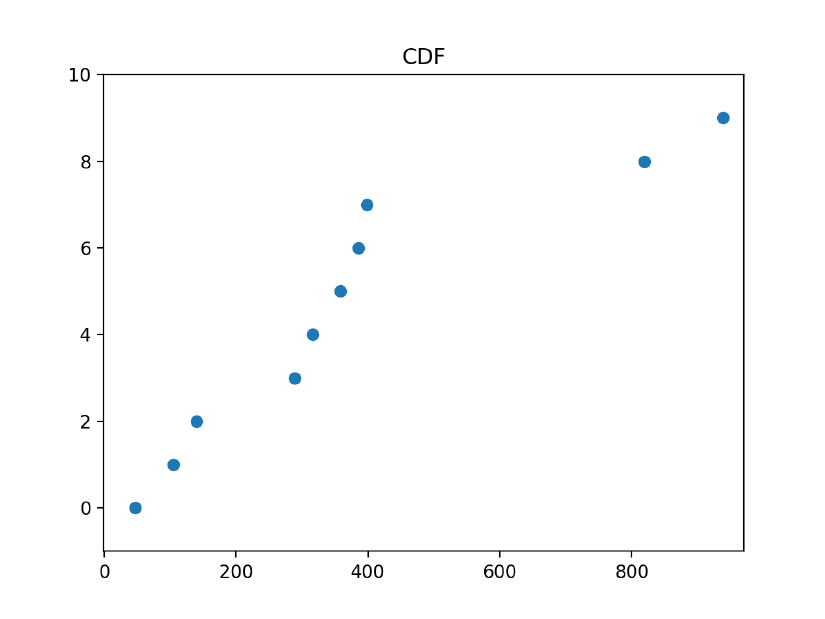}
		\end{minipage}\hfill
		(b)
		\begin{minipage}{0.25\textwidth}
			\includegraphics[width=\linewidth]{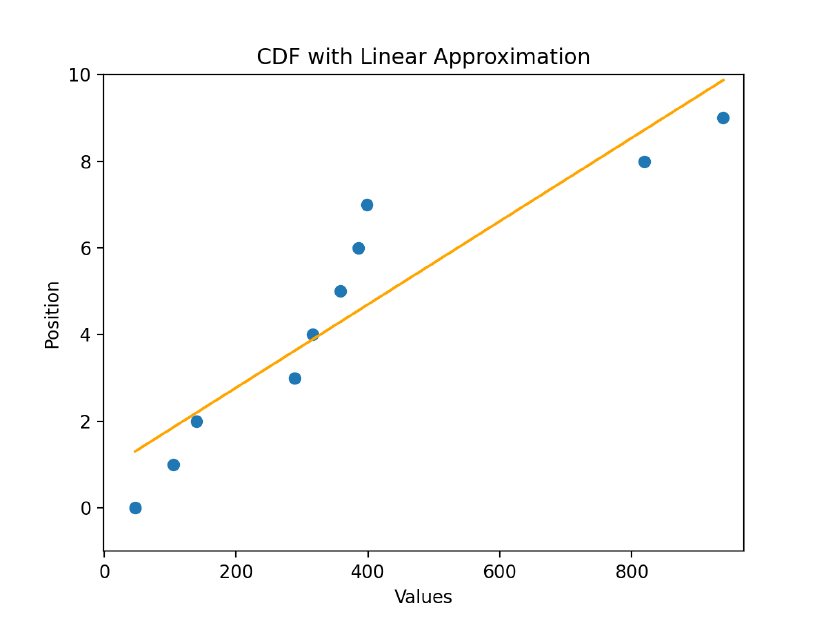}
		\end{minipage}\hfill
		(c)
		\begin{minipage}{0.25\textwidth}%
			\includegraphics[width=\linewidth]{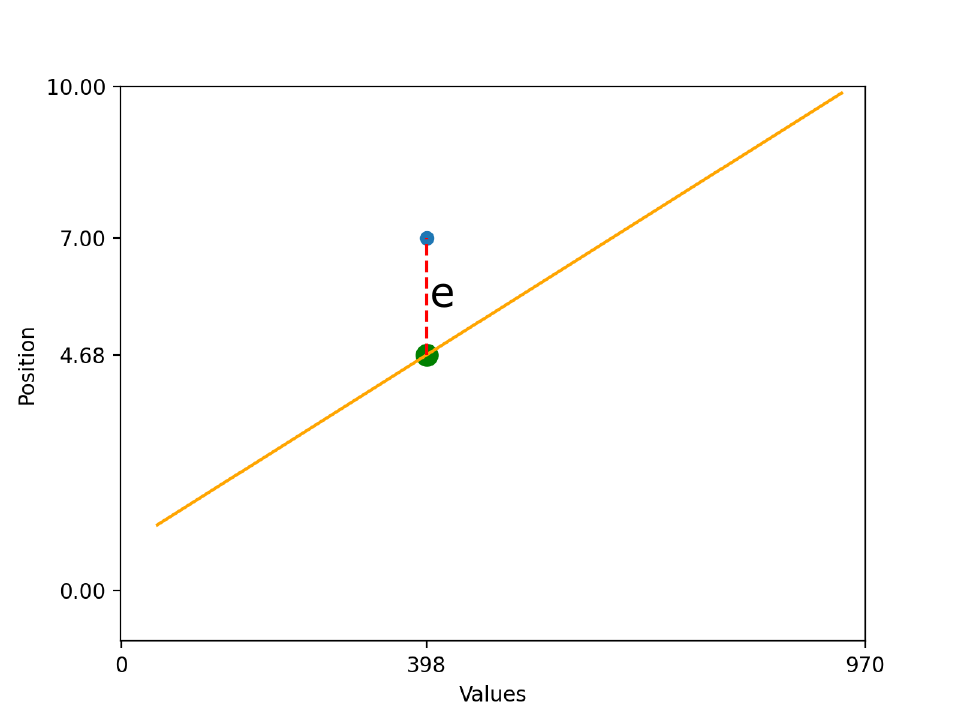}
		\end{minipage}
		\caption{{\bf  The Process of Learning a Simple Model via Linear Regression.} Let $A$ be $[47, 105, 140, 289, 316, 358, 386, 398, 819, 939]$. (a) The CDF of A. In  the diagram, the abscissa indicates the value of an element in the table, while the ordinate is its rank. (b) The straight line  $F(x)=ax+b$ is  obtained by determining $a$ and $b$ via Linear Regression, with Mean Square Error Minimization.  (c) The maximum error $\epsilon$ one can incur in using $F$ is also important. In this case, it is $\epsilon=3$, i.e., accounting for rounding,  it is the maximum distance between the rank of  a point in the table and its rank as predicted by $F$. In this case, the interval to search into, for a given query element $x$,  is given by $[F(x)-\epsilon, F(x)+\epsilon]$. }
		%\end{center}
		\label{fig:CDF}
	\end{figure}
	
	\begin{figure}[!htb]
		\centering
		\includegraphics[scale=0.8]{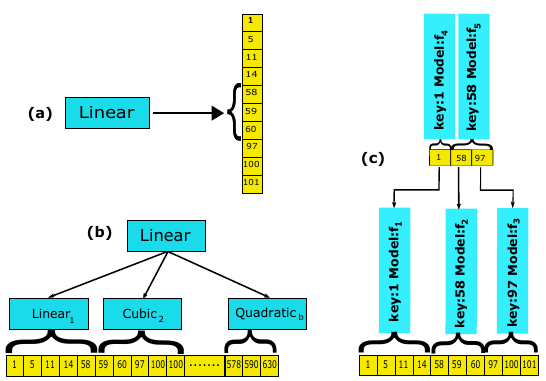}
		
		\caption{ {\bf  Examples of various Learned Indexes} (see also \cite{Marcus20}). (a) an Atomic Model, where the box linear means that the CDF of the entire dataset is estimated by a linear function via Regression, as exemplified in Figure \ref{fig:CDF}.  (b)  An example of an {\bf RMI} with two layers and a branching factor equal to $b$. The top box indicates that the lower models are selected via a linear function. As for the leaf boxes,  each indicates which Atomic  Model is used for prediction on the relevant portion of the table.   (c)  An example of a {\bf PGM }  Index. At the bottom, the table is divided into three parts. A new table is so constructed and the process is iterated.}  
		\label{fig:models}
	\end{figure}
	
	\section{A Simple Example of How to Build  and Query a Model}
	\label{sec:exquery}
	We now outline the basic technique that one can use to build a model for $A$. It relies on Linear Regression,  with Mean Square Error Minimization \cite{FreedmanStat}. 
	With reference to the example in Figure \ref{fig:CDF},  and assuming that one wants a linear model, i.e., $F(x)=ax+b$, Kraska et al. \cite{kraska18case}  note that they can fit a straight line to the Empirical Cumulative Distribution Function (CDF) and then use it to predict where a point $x$ may fall in terms of rank and accounting also for approximation errors. More in general, in order to perform a query, the model is consulted and an interval in which to search for is returned. Then, to fix ideas,  Binary Search on that interval is performed.

	\section{Models Specific for Binary and Interpolation Search}
	\label{sec:models}
	With reference to Figure  \ref{fig:models}, we provide a brief description of three models specific for Binary and Interpolation Search. That is, as of their current implementation, they work when $A$ is stored in a sorted table. 
	
	\begin{itemize}
		\item {\bf Atomic Models}. They are closed-form expressions that can be evaluated when queried, in order to get the predicted position in the table. An example is provided in Figure  \ref{fig:models}(a). Usually, the closed-form expression is a low degree polynomial approximating the CDF. The interested reader can find details in \cite{amato2021learned} on how such a polynomial can be derived from the table via Regression. 
		
		\item {\bf The RMI Family}. Informally, an {\bf RMI}  is a multi-level, directed graph, with Atomic Models at its nodes. When searching for a given key and starting with the first level, a prediction at each level identifies the model of the next level to use for the next prediction. This process continues until a final level model is reached. This latter is used to predict the table interval to search into. An  example  of {\bf RMI} is   provided in Figure \ref{fig:models}(b). As pointed out in \cite{Kipf19,Marcus20},  in most applications, a generic {\bf RMI} with two layers, a tree-like structure and a branching factor $b$ suffices to handle large tables. Amato \emph{et al.} \cite{amato2021learned} have shown that they perform well on a wide spectrum of table sizes.  It is to be noted that Atomic Models are {\bf RMI}s, although it is more convenient to single them out as \vir{basic building boxes} of more complex models. 
		
		\item  {\bf PGM}. It is also a multi-stage model, built bottom-up and queried top down.  It uses a user-defined approximation parameter $\epsilon$, that controls the prediction error at each stage. With reference to Figure \ref{fig:models}(c), the table is subdivided into three pieces. A prediction in each piece can be provided via a linear model guaranteeing an error of $\epsilon$.  A new table is formed by selecting the minimum values in each of the three pieces. This new table is possibly again partitioned into pieces, in which a linear model can make a prediction within the given error. The process is iterated until only one linear model suffices, as in the case in the Figure. A query is processed via a series of predictions, starting at the root of the tree. 
		
	\end{itemize}

	\section{Proof Outlines}
	\label{sec:proofs}
	The notation and the numbering of the Theorems are the same as in the main text. 
	\subsection{Theorem \ref{M-th:Dkconstruction}}
	The proof regarding the construction easily follows from the convexity of  $g(c)$ and the fact that it is non-decreasing. The proof regarding the search time is a simple extension of the proof of  Theorem  2.1. in \cite{Demaine04}.

	%\subsection{Theorem \ref{M-th:Dksearch}}
	%The proof  of  our Theorem  is a simple extension of  the proof  of  Theorem  2.1. in \cite{Demaine04}. 
	
	\subsection{Theorem \ref{M-th:smoothsearch}}
	Let $\mu_m$ be the expected number of elements in bin $m$. We need to prove that under the assumptions of the Theorem, the probability that $\mu_m > t\log^{O(1)} n$, $t$ constant, is bounded by $n^{-2}$. Then the result follows by the Union Bound. 
	
	To this end, we first show the following.  Assume that the elements in $A$ are drawn from $U$ according to some probability distribution and independently. Then, the  following events hold with 
	probability at most $n^{-2}$.   {\bf (A) The Lightly Loaded Case}.  When  $0< \mu_m< 2e \ln n$,  the number of elements in bin $m$ exceeds $2e^2\ln n$. 
	{\bf (B) The Heavily Loaded Case}. When $\mu_m\geq  2e \ln n$.,  the number of elements in that bin exceeds  $e\mu_m$. 
	
	In order to prove the above statements, we reduce the assignment of the elements to the $k$ bins to a Balls and Bins game, in which we are interested in the Maximum Expected Load of any bin. Then, the results follow by using standard Chernoff Bounds arguments \cite{Spirakis2008}.

	Now, using the assumption that the probability distribution from which the elements  in $A$ are drawn from $U$ is smooth, we show that $\mu_m \leq \beta \ln^{O(1)} n$.  Now, let $t=\max(\beta, 2e^2)$.  Observe that, for any bin $m$.

	$${\bf PR}[\text{num.  of elements in bin } m>t \ln^{O(1)} n] \leq n^{-2}.$$.
	
	Indeed, by (A), the result follows immediately for lightly loaded bins, while for heavily loaded bins, we need to use the bound on $u_m$ mentioned earlier and then (B). Now the result follows by the Union Bound, applied to all bins and from the fact that we have $f_1(n)=cn$ of them, for some constant $ 0< c \leq 1$.

	\subsection{Therem \ref{M-th:entropybounds}}
	We need to recall some facts from the Literature. 
	Given the probability distibution $P$ and $Q$, it is well known that one can build an Optimum Binary Search Tree \cite{Knuth71,Yao80}, i.e., one minimizing the expected search time $C(T)$, in $O(n^2)$ time \cite{Knuth71,Yao80}.  Moreover, $C(T_{opt}) \leq H(P,Q)+1$,  as shown in \cite{DEPRISCO93}, where it is also  provided  a linear time approximation algorithm 
	that produces a search tree  with cost bounded as in the optimum case in  terms of entropy.

	Based on the mentioned results, we now prove the Theorem.
	
	\emph{Part (a)}. Consider $D_k$ for a fixed  $k$, $1 \leq k \leq k_{max}$. For each bin in $D_k$, build an Optimum Binary Search Tree for the elements of $A$ falling in that bin, with the access probabilities relevant for that bin. 
	Let $T_{opt}$ be the Optimum  Binary Search Tree corresponding to $A$, with access probabilities $P$ and $Q$. Now, $C(T_{opt})+1$ is the cost of the Optimum   Binary Search Forest corresponding to $D_1$ and it is an upper bound to the cost of the Optimum  Binary Search Forest computed on all $k$'s. Using the mentioned entropy bound on optimum cost, one obtains the claimed bound. As for time, the Optimum Search Forest for $D_k$ can be computed in $O(\Sigma_{i=1}^{n} \ell_i^2)$ time, which is bounded by $O(n^2)$.  The claimed bound now follows.
	
	\emph{Part (b)}. It follows along the same lines as (a), except that now we use the approximation algorithm in \cite{DEPRISCO93}  as a \vir{subroutine}. In this case, for a single  $k$, the cost is bounded by $O(\Sigma_{i=1}^{n}\ell_i) =O(n)$.

	\subsection{Theorem \ref{M-th:dynamic}}

	We prove the following, as in  \cite{Demaine04}:
	
	\begin{itemize}
		\item[(1)] 	None of the $n/2$ update operations immediately after a rebuild  can add a value outside the range  $[A[1]- L\hat{\Delta}, A[n]+L\hat{\Delta}]$. 
		
		\item[(2)]	The maximum number of elements in a bin after $n$  update operations is $O(n/k)\hat{\Delta}^2)$. 
		
	\end{itemize}

	Now,  point (1) grants correctness. The worst-case bound comes from point  (2), the fact that each bin cannot contain more than $O(n)$ elements between two rebuild operations and the assumptions regarding the cost of each of the mentioned operations for  ${\cal SD}$ .

	As for rebuild operations, we notice that a rebuild can be done in $O(n)$ time when there are $n$ elements present in the structure.
	
	When a rebuild occurs because  $n/2$ updates have been performed, we charge the cost of rebuild to each element involved in those updates, $O(1)$ time each.  During the lifetime of the structure, each element in it is charged at most twice.
	
	When a rebuild occurs because of a change in $\Delta$, we observe the following. Let $\Delta_0, \cdots, \Delta_{max}$ be the changes in $\Delta$ that cause a rebuilt. We notice:
	(a) $C_{i+1}= \log \Delta_{i+1} -\log \Delta_i= \Omega(1)$, since   $\Delta_{i+1} \geq 2\Delta_i$ $0 \leq i<max$; (b) the sum of the $C_i$'s telescopes and it is bounded by $\log \Delta_{max}$.
	Now, when a rebuild occurs because of a change in $\Delta$, we charge $C_{i+1}$ to each element in the structure at the time of the change, assuming that this is the $i$-th change. 
	This is enough to pay for the rebuild, by (a). 
	Because of (b), during the lifetime of the structure, an element can be charged at most $\log \Delta_{max}$ units.
	
	The bound on rebuild operations follows.

	\section{ The Role of $\Delta$}	
	
	In order to assess the role of $\Delta$ on real datasets, we proceed as follows. From the real datasets used in this research, we produce smaller ones that have the same statistical properties as the original set, i.e., CDF and PDF.
	Then, we compute the values of $\Delta$ for all the datasets we have either available or that we have generated.

	With reference to the datasets mentioned in Section \ref{M-sec:expm}  of the Main Text, we produce sorted sets of varying sizes, so that each fits in a level of the internal memory hierarchy, as follows. 
	Letting $n$ be the number of elements in a set, for the computer architecture we are using (Section \ref{M-sec:expm} of the Main text), the details of the tables we generate are as follows.

	\begin{enumerate}
		
		\item [$\bullet$] {\bf Fitting in L1 cache: cache size 64Kb.}  Therefore, we choose $n=3.7K $. For each dataset, the table corresponding to this type is denoted with the prefix {\bf L1}, e.g., {\bf L1\_amzn}, when needed.  
		For each dataset, in order to obtain a CDF that resembles one of the original tables, we proceed as follows. Concentrating on {\bf amzn}, since for the other datasets the procedure is analogous, we extract uniformly and at random a sample of the data of the required size. For each sample, we compute its CDF. Then, we use the Kolmogorov-Smirnov test \cite{smirnov1939estimate} in order to assess whether the CDF of the sample is different from the {\bf amzn32} CDF. 
		
		If the test returns that we cannot exclude such a possibility, we compute the PDF of the sample and compute its KL divergence \cite{kullback1968information} from the PDF of {\bf amzn}. We repeat such a process 100 times and, for our experiments, we use the sample dataset with the smallest KL divergence.

		\item [$\bullet$] {\bf Fitting in L2 cache: cache size 256Kb.} Therefore, we choose $n= 31.5K$. For each dataset, the table corresponding to this type is denoted with the prefix {\bf L2}, when needed. For each dataset, the generation procedure is the same as the one of the {\bf L1} dataset.

		\item [$\bullet$] {\bf Fitting in L3 cache: cache size 8Mb.} Therefore, we choose $n=750K$. For each dataset, the table corresponding to this type is denoted with the prefix {\bf L3}, when needed.  For each dataset, the generation procedure is the same as the one of the {\bf L1} dataset.

	\end{enumerate}
	For completeness, the results of the Kolmogorov-Smirnov Test as well as KL divergence computation are reported in Table \ref{T:ks}. For each memory level (first row) and each original dataset (first column), it is reported the percentage of times in which the Kolmogorov-Smirnov test failed to report a difference between the CDFs of the dataset extracted uniformly and at random from the original one and this latter, over 100 extractions. Moreover, the KL divergence between the PDFs of the chosen generated dataset and the original one is also reported. From those results, it is evident that the PDF of the original datasets is quite close to the PDF of the extracted datasets.

	\begin{table}
		\begin{center}
			\caption{ The results of the Kolmogorov-Smirnov Test and of the KL divergence  computation.}\label{T:ks}
			\begin{tabular}{|c|c|c|c|c|c|c|}
				\hline 
				& \multicolumn{2}{c}{{\bf L1}} \vline & \multicolumn{2}{c}{{\bf L2}} \vline & \multicolumn{2}{c}{{\bf L3}} \vline\\
				\hline 
				Datasets &\%succ &KLdiv  &\%succ &KLdiv &\%succ &KLdiv\\
				%\hline
				%{\bf amzn32} &100 &1.87e-05$\pm$1.41e-13 &100 &1.58e-04 $\pm$8.76e-13 &100 &3.77e-03 $\pm$2.20e-11 \\ 
				\hline
				{\bf amzn} &100 &9.54e-06$\pm$7.27e-14 &100 &7.88e-05$\pm$7.97e-13 &100 &1.88e-03$\pm$1.52e-11 \\ 
				\hline
				{\bf face} &100 &1.98e-05$\pm$1.00e-12 &100 &7.98e-05$\pm$4.43e-13 &100 &1.88e-03$\pm$1.24e-11 \\ 
				\hline
				{\bf osm} &100 &9.38e-06$\pm$4.51e-14 &100 &7.88e-05$\pm$3.46e-13 &100 &1.88e-03$\pm$9.55e-12 \\ 
				\hline
				{\bf wiki} &100 &9.47e-06$\pm$5.27e-14 &100 &7.87e-05$\pm$5.64e-13 &100 &1.88e-03$\pm$1.25e-11 \\ 
				\hline
			\end{tabular}
		\end{center}
	\end{table}

\begin{figure}
		\centering
		
		\includegraphics[scale=0.5]{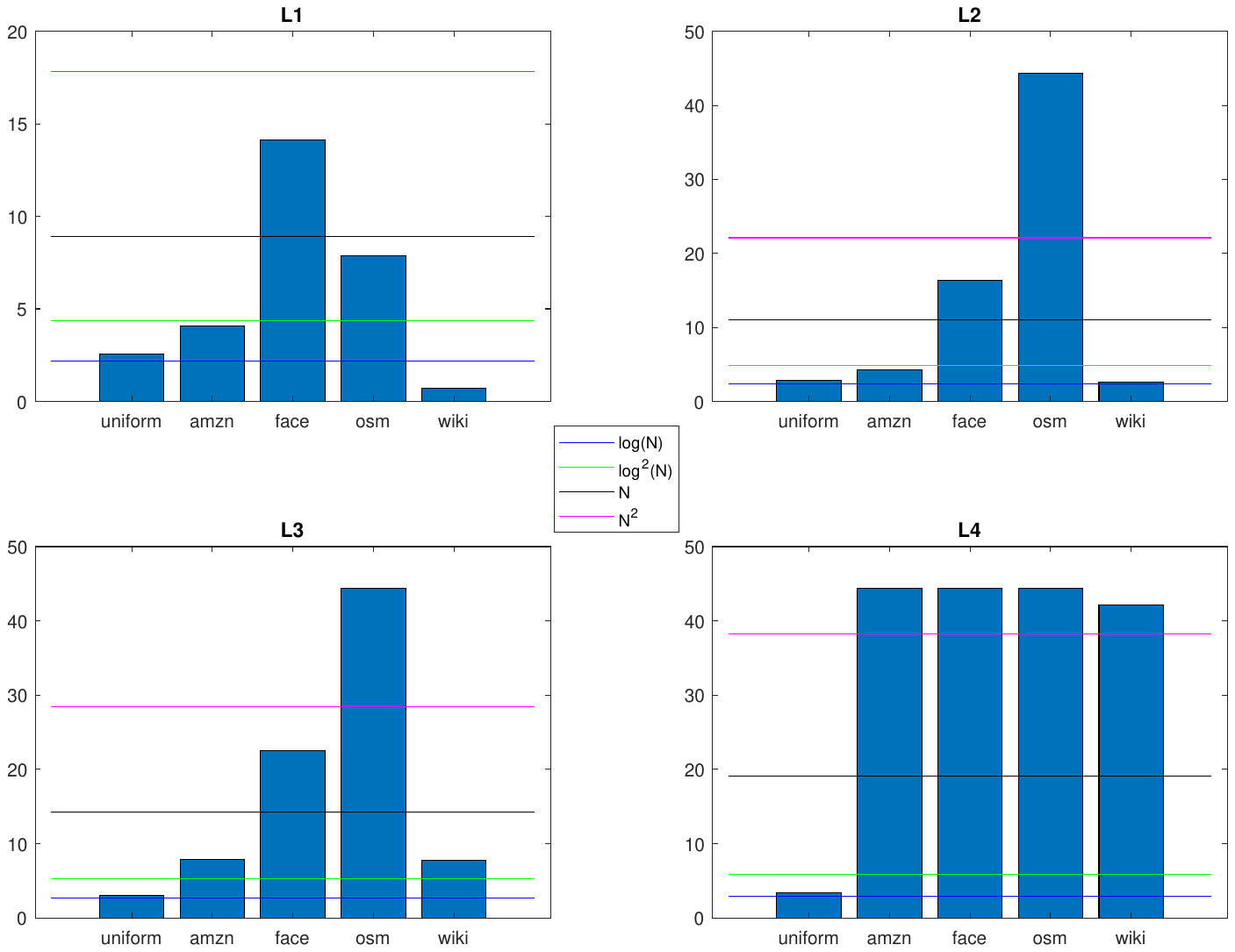}
		\caption{{\bf The Role of $\Delta$}.  To the collection of datasets discussed in the text of this Supplementary File, we have also added datasets generated from the Uniform Distribution, as a baseline. In each Figure, the horizontal lines denote various values of functions of the input size, which is the number of elements in each set. The abscissa contains the names of the datasets and the ordinate is the value of $\Delta $ for that dataset. Finally, the label on top of each Figure indicates the size of the dataset, as discussed in the text of this Supplementary File.} \label{img:L4delta}
	\end{figure}

	Figure \ref{img:L4delta} report the full set of experiments regarding the role of $\Delta$ on real datasets, briefly discussed in Section \ref{M-sec:delta} of the Main text.

	\section{Experiments: Results and Discussion-Additional Figures}
	\label{sec:experiments}
	Experiments have been carried out on Intel Core i7-8700 3.2GHz CPU with three levels of cache memory: (a) 64kb of L1 cache; (b) 256kb of L2 cache; (c)12Mb of shared L3 cache. The total amount of system memory is 32 Gbyte of DDR4. The operating system is Ubuntu LTS 20.04.

	\

	\begin{figure}[tbh]
		%\begin{center}
		\centering
		\includegraphics[scale=0.5]{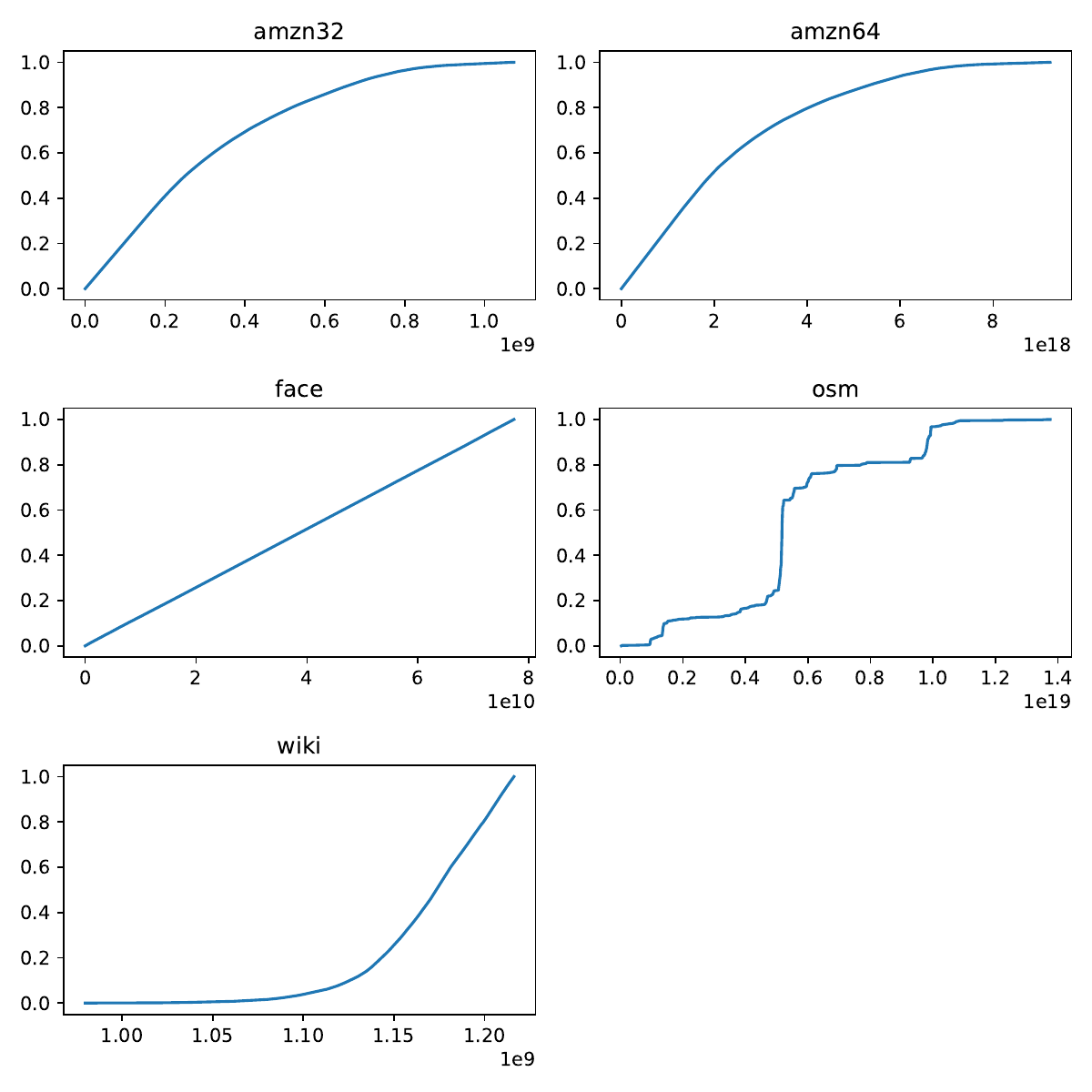}
		\caption{{ \bf  CDF plots}. 	 The Empirical
			Cumulative Distribution Fuction of each  dataset considered in this research.}
		%\end{center}
		\label{fig:cdf_plots}
	\end{figure}

	\begin{figure}[tbh]
		%\begin{center}
		\centering
		\includegraphics[scale=0.5]{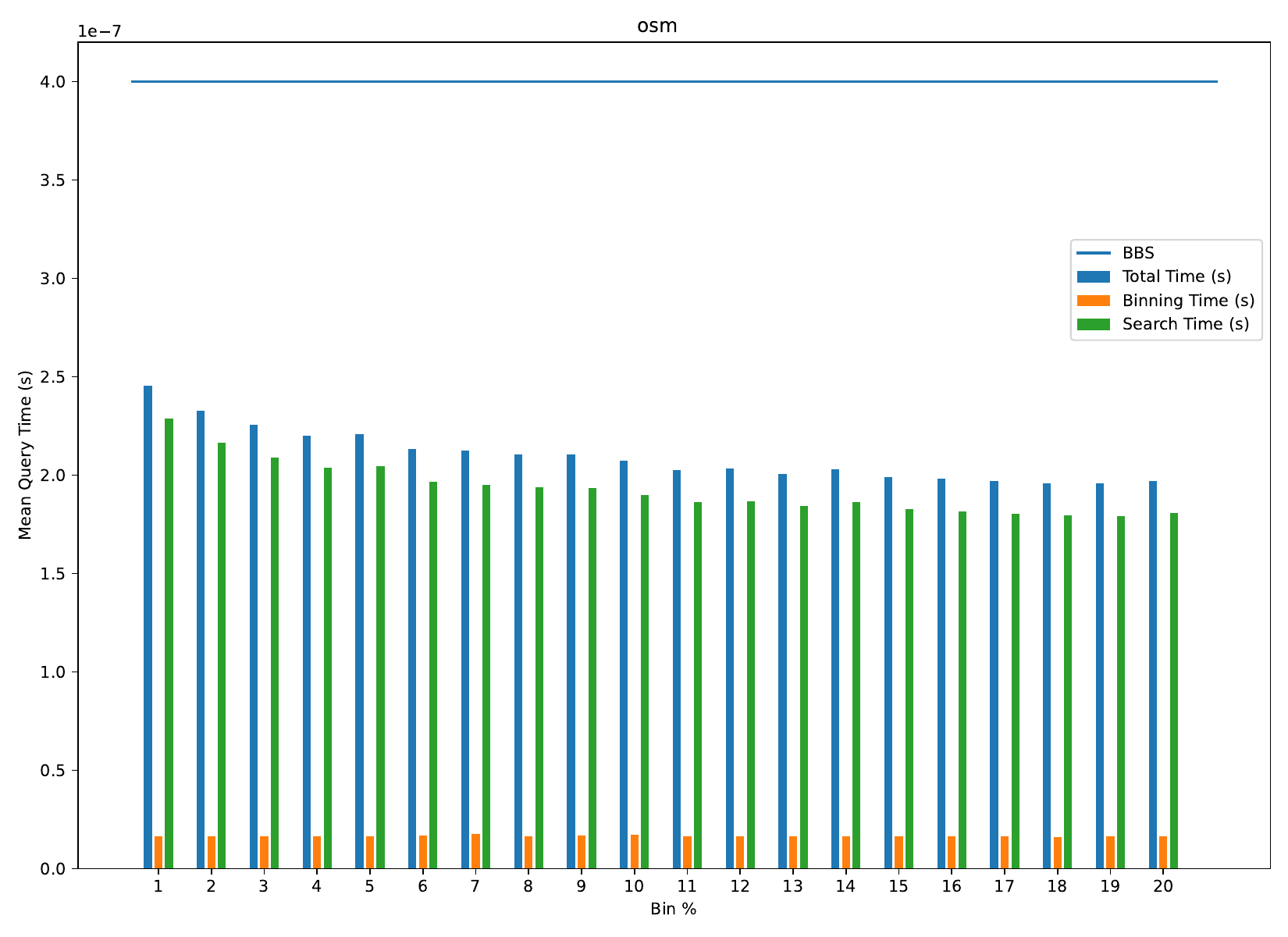}
		\caption{ { \bf Binning Query Time on the osm dataset}. We report on the $x$ axis the chosen percentage for the Binning construction (see Main Text), while on the $y$ axis the mean query time is expressed in seconds. The blue bars indicate the mean total time needed to execute a query using the Binning Model. The orange bar shows the mean time needed to make a prediction with the Binnig Model. The green bar reports the time to search in the predicted interval using {\bf BBS}. Finally, the horizontal line is the stand-alone {\bf BBS} mean query time. }
		%\end{center}
		\label{fig:bin_time}
	\end{figure}
	
	\begin{figure}[tbh]
		%\begin{center}
		\centering
		\includegraphics[scale=0.5]{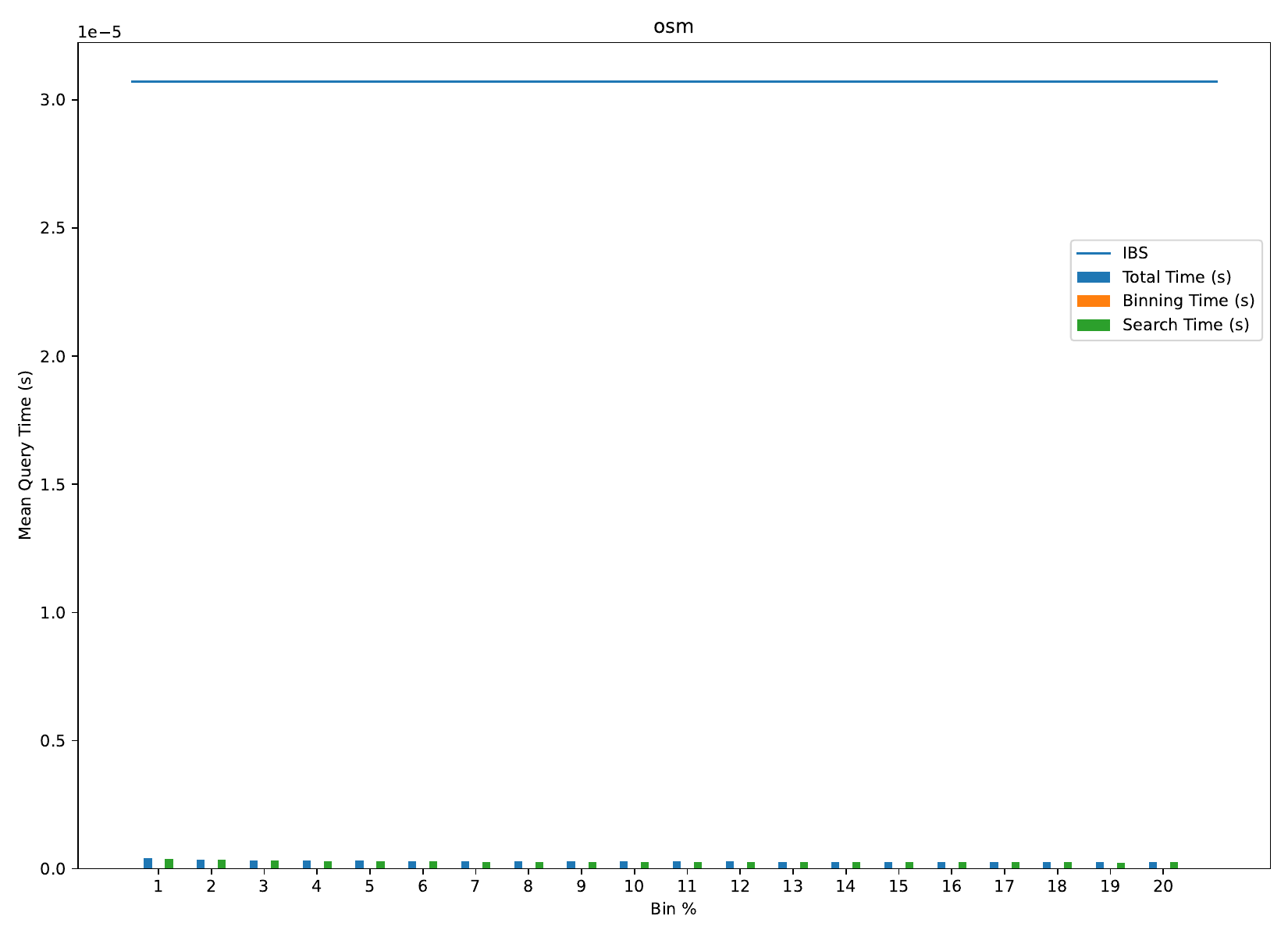}
		\caption{ { \bf Binning Query Time on the osm dataset}. We report on  the $x$ axis the chosen percentage for the Binning construction (see Main Text), while on the $y$ axis the mean query time is expressed in seconds. The blue bars indicate the mean total time needed to execute a query using the Binning Model. The orange bar shows the mean time needed to make a prediction with the Binnig Model. The green bar reports the time to search in the predicted interval using {\bf IBS}. Finally, the horizontal line is the stand-alone {\bf IBS} mean query time. }
		%\end{center}
		\label{fig:bin_time_ibs}
	\end{figure}
	
	\begin{figure}[tbh]
		%\begin{center}
		\centering
		\includegraphics[scale=0.5]{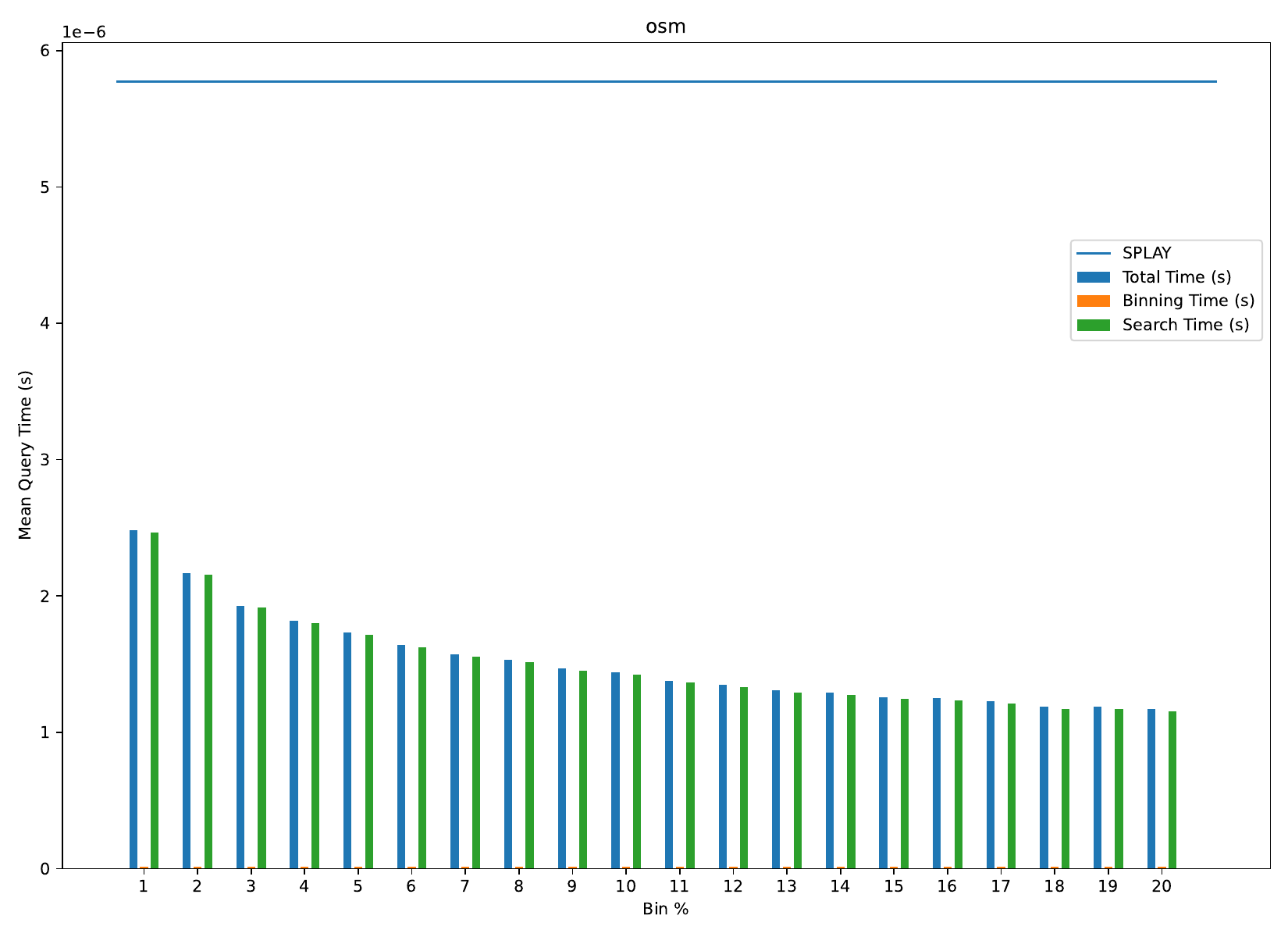}
		\caption{ { \bf Binning Query Time on the osm dataset}. We report on the $x$ axis the chosen percentage for the Binning construction (see Main Text), while on the $y$ axis the mean query time is expressed in seconds. The blue bars indicate the mean total time needed to execute a query using the Binning Model. The orange bar shows the mean time needed to make a prediction with the Binnig Model. The green bar reports the time to search in the predicted interval using {\bf SPLAY}. Finally, the horizontal line is the stand-alone {\bf SPLAY} mean query time. }
		%\end{center}
		\label{fig:bin_time_splay}
	\end{figure}

	\begin{figure}[tbh]
		%\begin{center}
		\centering
		\includegraphics[scale=0.5]{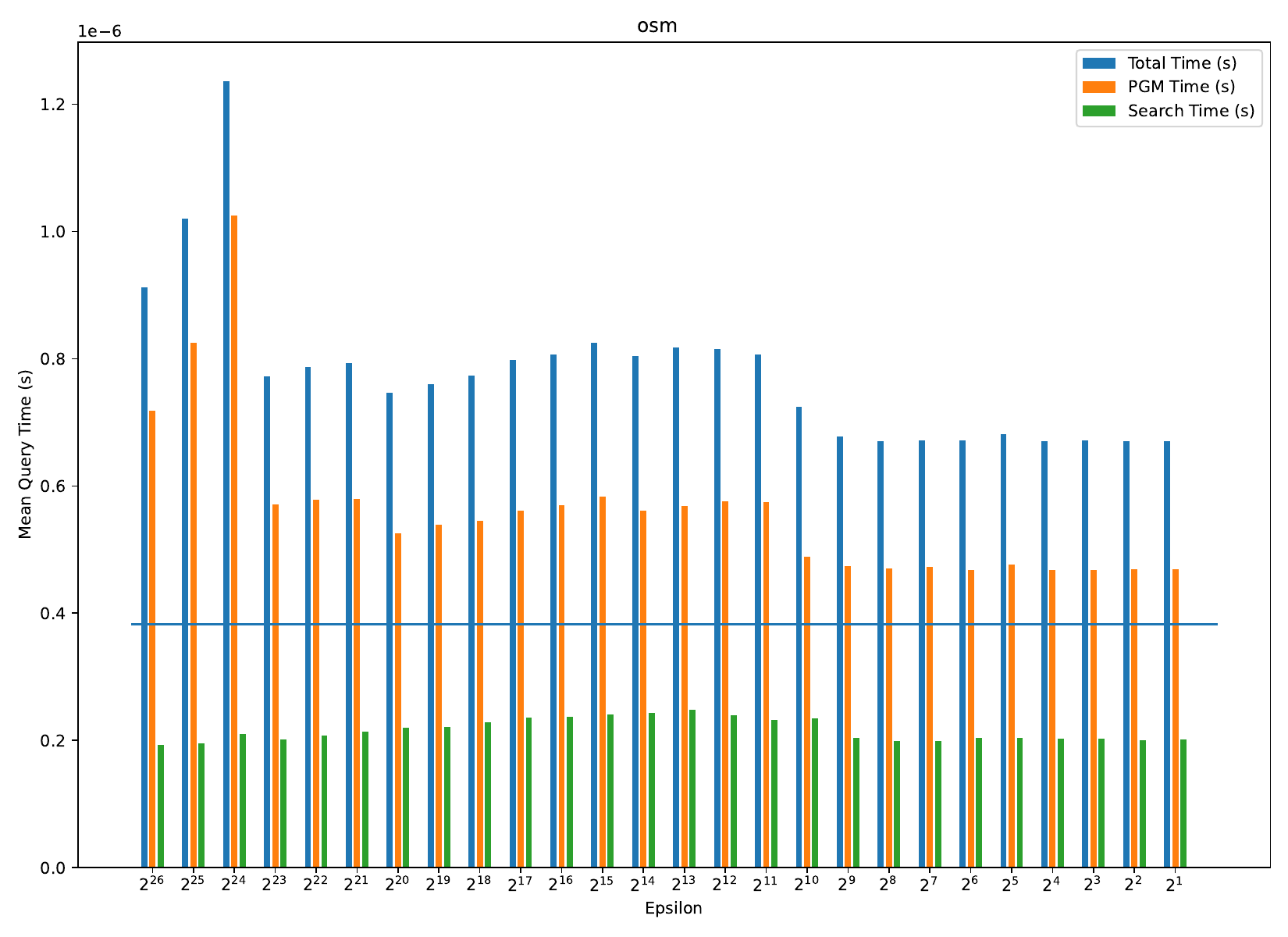}
		\caption{{\bf PGM Query Time on the osm dataset}. We report on the $x$ axis the chosen $\epsilon$ for the PGM construction, while on the $y$ axis the mean query time is expressed in seconds. The blue bars indicate the mean total time to execute a query using the PGM Model. The orange bar shows the mean time taken to navigate the PGM Model, in order to get a prediction. The green bar reports the mean time to search in the predicted interval using  {\bf BBS}. Finally, the blue horizontal line is the stand-alone {\bf BBS}  mean query time.  }
		%\end{center}
		\label{fig:pgm_time}
	\end{figure}
	
	\begin{figure}[tbh]
		%\begin{center}
		\centering
		\includegraphics[scale=0.5]{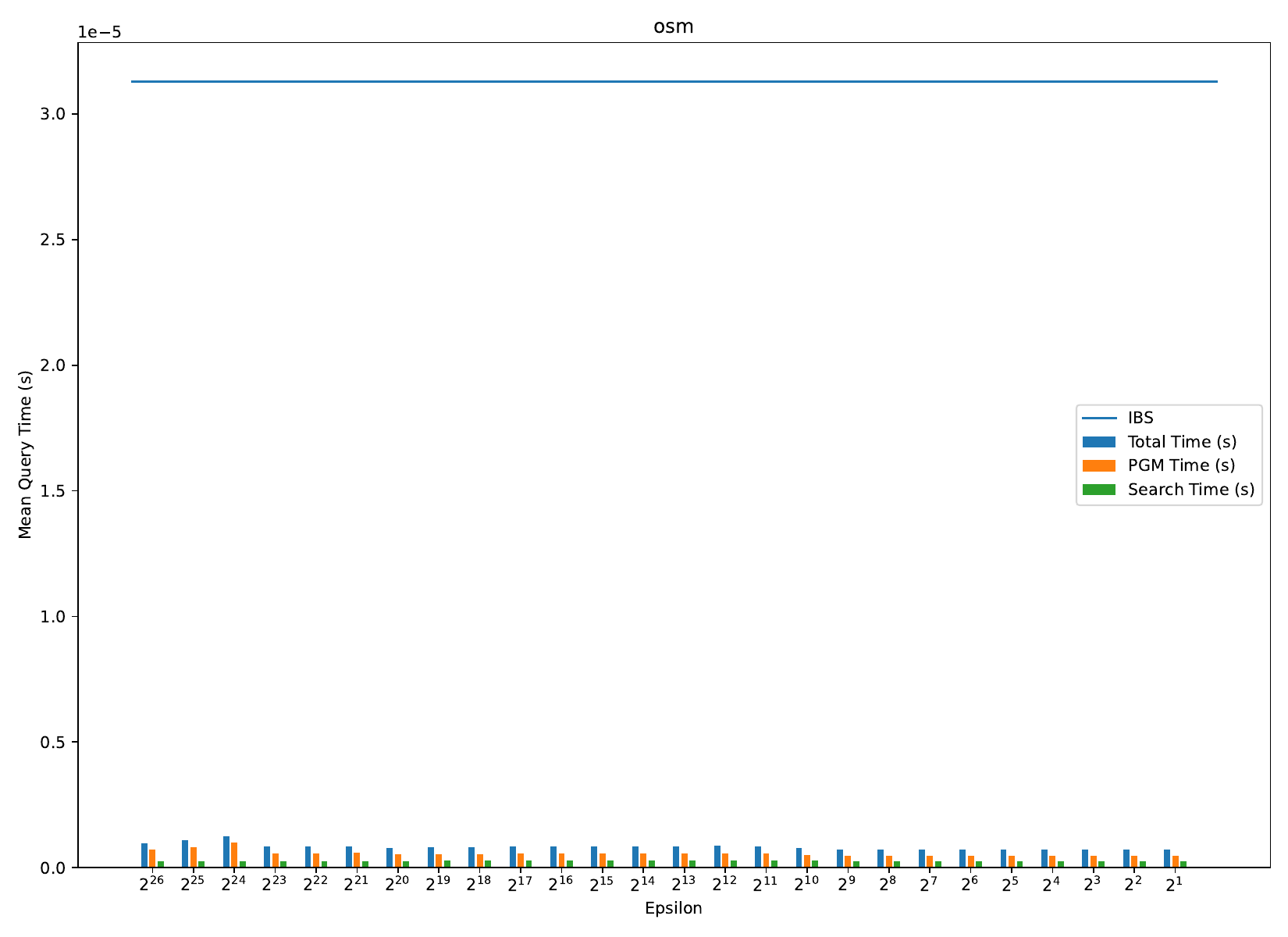}
		\caption{{\bf PGM Query Time on the osm dataset}. We report on the $x$ axis the chosen $\epsilon$ for the PGM construction, while on the $y$ axis the mean query time is expressed in seconds. The blue bars indicate the mean total time to execute a query using the PGM Model. The orange bar shows the mean time taken to navigate the PGM Model, in order to get a prediction. The green bar reports the mean time to search in the predicted interval using  {\bf IBS}. Finally, the blue horizontal line is the stand-alone {\bf IBS}  mean query time.  }
		%\end{center}
		\label{fig:pgm_time_ibs}
	\end{figure}
	
	\begin{figure}[tbh]
		%\begin{center}
		\centering
		\includegraphics[scale=0.5]{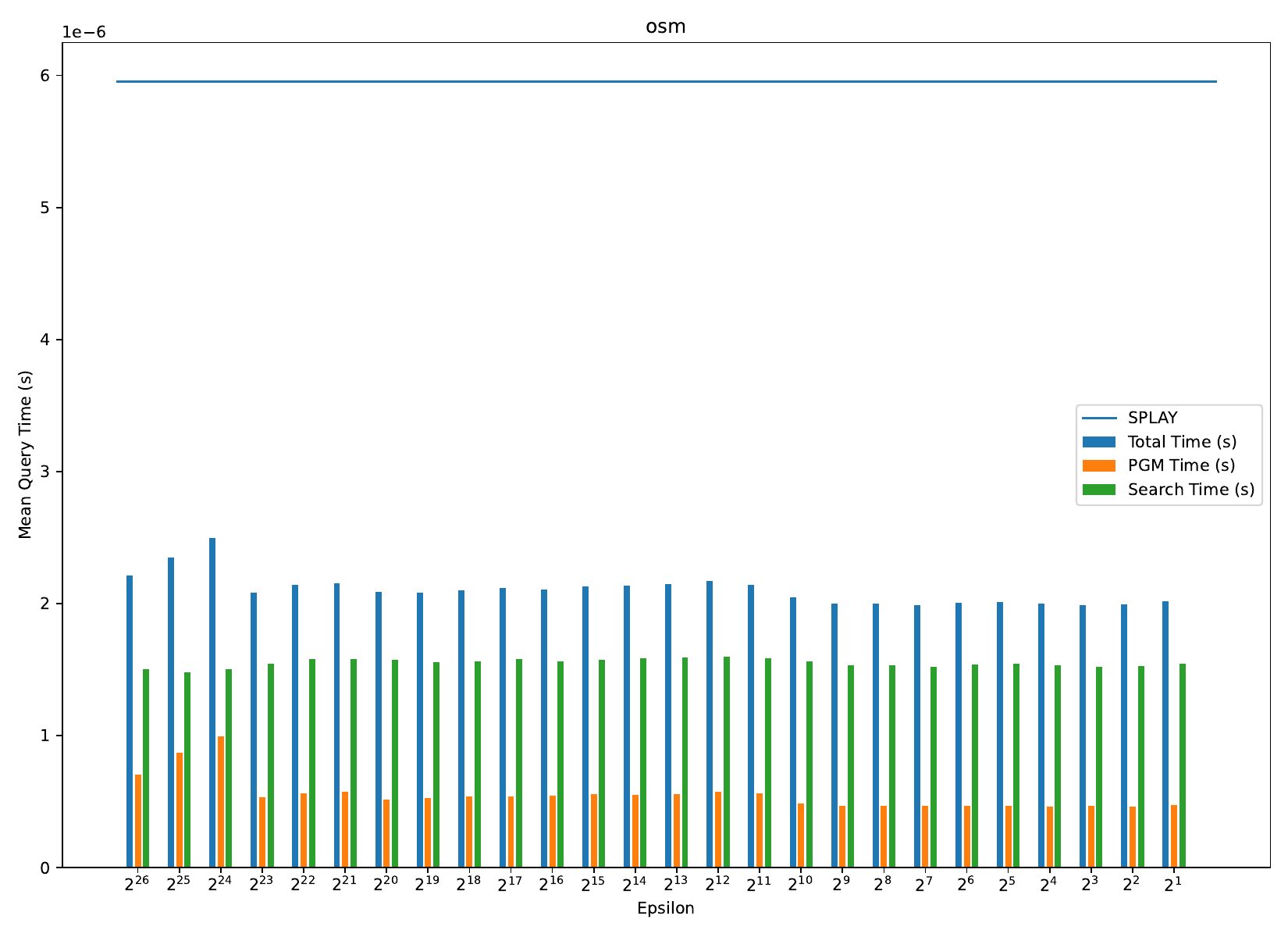}
		\caption{{\bf PGM Query Time on the osm dataset}. We report on the $x$ axis the chosen $\epsilon$ for the PGM construction, while on the $y$ axis the mean query time is expressed in seconds. The blue bars indicate the mean total time to execute a query using the PGM Model. The orange bar shows the mean time taken to navigate the PGM Model, in order to get a prediction. The green bar reports the mean time to search in the predicted interval using {\bf SPLAY}. Finally, the blue horizontal line is the stand-alone  {\bf SPLAY }  mean query time.  }
		%\end{center}
		\label{fig:pgm_time_splay}
	\end{figure}
	
	\clearpage
		
	\bibliographystyle{plain}
	\bibliography{references}